\documentclass{SciPost}

\usepackage{graphicx}
\usepackage{dcolumn}
\usepackage{physics}
\usepackage{subcaption}
\usepackage{xcolor}
\usepackage{mathtools}
\usepackage{bm}
\usepackage{cite}
\usepackage{doi}
\usepackage{tikz-cd}

\usepackage{upgreek}  
\usepackage{import}  

\newcommand{\RR}{\mathbb{R}}
\newcommand{\ZZ}{\mathbb{Z}}
\newcommand{\TT}{\mathbb{T}}
\newcommand{\Ac}{\mathcal{A}}
\newcommand{\Fc}{\mathcal{F}}
\newcommand{\Ham}{\mathcal{H}}
\newcommand{\KS}{K^2\times S^1}
\newcommand{\ZZtwist}{\widetilde{\ZZ}}
\newcommand{\bk}{\mathbf{k}}
\newcommand{\bd}{\mathbf{d}}

\newcommand{\noncoloneqq}{\phantom{\vcentcolon\mathrel{\mkern-1.2mu}}=}  

\renewcommand{\setminus}{-}  

\DeclareMathOperator{\im}{im}
\DeclareMathOperator{\Real}{Re}
\DeclareMathOperator{\Imag}{Im}

\hypersetup{
    colorlinks,
    linkcolor={red!50!black},
    citecolor={blue!50!black},
    urlcolor={blue!80!black}
}

\usepackage[bitstream-charter]{mathdesign}
\DeclareSymbolFont{usualmathcal}{OMS}{cmsy}{m}{n}
\DeclareSymbolFontAlphabet{\mathcal}{usualmathcal}

\fancypagestyle{SPstyle}{
\fancyhf{}
\lhead{\colorbox{scipostblue}{\bf \color{white} ~SciPost Physics }}
\rhead{{\bf \color{scipostdeepblue} ~Submission }}

\fancyfoot[C]{\textbf{\thepage}}
}

\begin{document}

\pagestyle{SPstyle}

\begin{center}{\Large \textbf{\color{scipostdeepblue}{
Twisted (co)homology of non-orientable Weyl semimetals
}}}\end{center}

\begin{center}\textbf{
Thijs Douwes\textsuperscript{1}
and
Marcus St{\aa}lhammar\textsuperscript{1,2$\star$}
}\end{center}

\begin{center}
{\bf 1} Institute of Theoretical Physics, Utrecht University \\
    Princetonplein 5, 3584CC Utrecht, The Netherlands
\\
{\bf 2} Department of Physics and Astronomy, Uppsala University \\
Box 524, 75120, Uppsala, Sweden
\\[\baselineskip]
$\star$ \href{mailto:email1}{\small marcus.backlund@physics.uu.se}
\end{center}

\section*{\color{scipostdeepblue}{Abstract}}
\textbf{\boldmath{The quasi-particle excitations in Weyl semimetals, known as Weyl fermions, are usually forced to emerge in charge-conjugate pairs by the Nielsen--Ninomiya theorem.
When the Brillouin zone is non-orientable, this constraint is replaced by a $\ZZ_2$ charge cancellation, as a result of the chirality becoming ill-defined on such manifolds; this results in configurations with seemingly non-zero total chirality.
Here, we set out to explain this behaviour from a purely topological perspective, and provide a classification of non-orientable Weyl semimetal topology in terms of exact sequences of twisted (co)homology groups.
This leads to several discoveries of direct physical importance: in particular, we recover the $\ZZ_2$ charge cancellation in a coordinate-independent way, allowing meaningful limits to be set on its physical interpretation. 
A detailed discussion is provided on a specific Klein bottle-like topology induced by a momentum-space glide symmetry, including a full review of the insulating and semimetallic invariants of the system and a classification of the surface states on the non-orientable boundary.
Beyond this, we survey all possible non-orientable Brillouin zones and their semimetallic invariants, and extend our formalism into the realm of non-Hermitian topological physics and inversion-symmetric Weyl semimetals.
Our work exemplifies the potential of fundamental mathematical descriptions to aid the physical intuition, predict novel and hitherto overlooked phenomena, and find concrete applications in areas such as acoustics and photonics.}}

\vspace{\baselineskip}



\vspace{10pt}
\noindent\rule{\textwidth}{1pt}
\tableofcontents
\noindent\rule{\textwidth}{1pt}
\vspace{10pt}

\section{Introduction}
\label{sec:intro}
The fusion of the mathematical framework of topology and condensed matter physics has revolutionised the understanding of quantum matter during the past decades.
Following the hallmark discovery of the quantum Hall effect in the 1980s~\cite{Klitzing1980}, a prominent advancement was realised in the classification of Bloch bands~\cite{Budich2013}, paving the way towards a more rigorous understanding of the topological phases of matter. These phases include gapped insulators and superconductors~\cite{hasankane,qizhang}, as well as gapless semimetals such as graphene~\cite{goerbig} and Weyl semimetals~\cite{weylreview,HQ2013}.
The latter are three-dimensional materials hosting so-called Weyl points or Weyl nodes, where the valence and conduction bands coincide. These band crossing points are considered accidental, in the sense that they are not enforced by any symmetry of the system.
They are, however, topological in nature, which ensures that they are robust to external perturbations; for example, a Weyl semimetal may be subjected to some disorder without removing the Weyl nodes.
Weyl semimetals find profound applications not only in material science~\cite{XBAN2015,LWFM2015,WSMTaP}, where their transport properties suggest technical applications in future solar cells~\cite{Nagaosa2017,Liu2020,Chen2024}, but also in particle physics, allowing Weyl fermions to be studied as quasi-particle excitations~\cite{W1929}, in photonic crystals~\cite{LWYRFJS2015}, and in constructing laboratory setups which potentially mimic features in analogue quantum gravity such as black holes~\cite{Volovik2016,zobkovblackholes,volovik,Zubkovold,Liu2018,Yaron2020,Nissinen}.

From a physical point of view, the topological properties of Weyl nodes are usually thought of in terms of quasiparticle charges: each Weyl node can be assigned an integer monopole charge describing the flux of a gauge field around the corresponding quasiparticle~\cite{weylreview,HQ2013}. 
These charges are explicitly chiral in nature, relating to the chirality of the corresponding Weyl fermion quasiparticles.
In periodic structures, where the Brillouin zone takes the form of a three-dimensional torus, the total charge is constrained to vanish, forbidding any non-zero net gauge flux~\cite{ZXBY2016,XPYQ2015,SY2012,G2012,ZB2012,GT2013,PGAPV2014,KGM2015,KGM2017,BKS2018,Huangetal2015}.
This constraint is known as the Nielsen--Ninomiya theorem, and it effectively forces the Weyl nodes to emerge in pairs~\cite{NM1981a,NM1981b,NM1983}, with each node being associated with a partner of opposite chirality.
Upon merging within the Brillouin zone, positive and negative Weyl nodes may annihilate in pairs, leaving a gapped band structure.

Mathematically, these features can be formally described using the theory of characteristic classes and the language of homology and cohomology~\cite{Mathai2017a,Mathai2017b}.
Studies have shown that relatively simple arguments involving a so-called Mayer--Vietoris sequence of cohomology groups can provide a deeper understanding of the structure and origin of these topological features of Weyl semimetals.
For example, the Nielsen--Ninomiya theorem follows directly from the exactness of this sequence; the well-established Poincar\'e duality between homology and cohomology groups generates an interpretation of Weyl point topology in terms of \emph{Dirac strings} (strings of gauge singularities between the Weyl nodes); the celebrated bulk--boundary correspondence is formalised by letting the surface projection act on homology groups, providing mathematical context to surface Fermi arcs as projections of these Dirac strings; and finally, the entries of the (co)homology groups classify the possible topological invariants, corresponding precisely to topologically distinct Berry curvatures.
The (co)homology groups can hence be used to obtain a relatively straightforward, yet formally correct, understanding of the global and local topology of Weyl semimetals, without relying on far more involved classifications based on K-theory~\cite{Altland1997, kitaev_periodic_2009, ryu_topological_2010, schnyder_classification_2008,Freed2013,Thiang2015,Gomi2015,Shiozaki2022}.
Complementary to the above are homotopy classifications, where the object of interest is the arguably more abstract classification space of the corresponding Hamiltonian~\cite{Bzdusek2017,Sun2018a}; these provide a direct route towards generalisation to an arbitrary number of bands, while the physical impact and interpretation of topological invariants remain somewhat elusive.

Although Weyl nodes do not rely on additional symmetries to arise, the presence of such symmetries may in fact alter their topological properties, and hence their (co)homology description.
An experimentally significant example of this is the impact of time reversal symmetry.
To account for this symmetry, the authors of Ref.~\cite{Thiang2017} show that the (co)homology sequences, as well as the conventional Poincar\'e duality, need to be modified.
The correct classification is then performed using so-called equivariant homology and twisted equivariant cohomology groups~\cite{Thiang2017,Nittis2018}, and includes a complete description of all (weak and strong) invariants for gapped (insulating) as well as gapless (semimetallic) phases.
The presence of time reversal symmetry effectively halves the Brillouin zone into \emph{fundamental domains}; all physical properties of momenta in one half are mirrored to the other half, with the exception of the eight time-reversal invariant momenta; the presence of these special momenta prevents the fundamental domain from being considered a Brillouin zone in its own right.

Recently, a class of non-symmorphic (i.e.\ combining rotation/reflection with fractional lattice translations) momentum-space symmetries that lack such fixed momenta has been studied, leading to fundamentally altered Brillouin zone topologies. 
Of particular interest in this work are the orientation-reversing symmetries, under which the fundamental domain becomes non-orientable~\cite{fonseca2024,Chen2022,Tao2024,Zhu2024,Wang2023}.
This loss of orientability causes the chirality of Weyl nodes to become ill defined; this has recently been used to study setups where the constraints imposed by the Nielsen--Ninomiya theorem are circumvented.
Concretely, the authors of Ref.~\cite{fonseca2024} show that Weyl nodes emerging in non-orientable Brillouin zones are not necessarily charge-conjugate to each other: a Weyl node with topological invariant $\pm 1$ is no longer always accompanied by a partner with topological invariant $\mp 1$; instead, the partner may have the same topological invariant.
In general terms, the authors derive a mod 2 charge cancellation theorem, rather than the conventional complete charge cancellation imposed by the Nielsen--Ninomiya theorem.
Similar mod 2 charge cancellation rules for Weyl points have been observed within the regime of non-Hermitian physics, where gain and loss features may induce a chirality flip~\cite{Wojcik2020,Sun2020}, much like those induced by so-called Alice strings~\cite{Schwarz1982}.

Although the work in Ref.~\cite{fonseca2024} is of indisputable importance, highlighting the vital role that the topology of the Brillouin zone plays in Weyl semimetal topology, the arguments given have a crucial limitation: the definition of the topological invariants depends on a particular choice of a fundamental domain, which amounts to a choice of orientation upon integration.
This ambiguity highlights the imminent need for a ``coordinate-free'' topological description, in the sense that it is independent of any particular position of the fundamental domain.
We develop such a description in this work, applicable to systems in which non-orientable fundamental domains are induced by symmetries, and use the particular system studied in Ref.~\cite{fonseca2024} as an illustrative example.
We modify the aforementioned classification scheme in terms of (co)homology groups, showing that the homology groups must be twisted in order to account for the loss of orientability under this specific symmetry. The use of these twisted homology groups gives rise to the concepts of \emph{twisted Dirac strings} and \emph{twisted Fermi arcs}, which can be used to great effect to gain intuition for the underlying topological invariants and their physical interpretation.
We also directly verify the mod 2 charge cancellation found in Ref.~\cite{fonseca2024}, now within a mathematically rigorous and coordinate-free context. In addition, we discuss an emergent $\ZZ$ invariant associated with this specific system.

To illustrate the utility of this more abstract classification scheme, we further investigate all other non-symmorphic momentum-space symmetries which give rise to non-orientable Brillouin zones, providing the associated groups of invariants in each case.
We also show how this description can be applied to non-Hermitian systems to describe the topology of exceptional points (non-Hermitian degeneracies where not only the energy bands but also the associated eigenstates coalesce) in non-orientable fundamental domains, formalising the two-band case of the very recent works in Refs.~\cite{Konig2025,Rui2025}.
Lastly, we also provide heuristic arguments addressing the highly relevant case of inversion-symmetric Weyl semimetals within this framework, further exemplifying its usefulness and relevance.
Our work thus provides a fundamental mathematical understanding of the topology of gapped and gapless systems with non-orientable fundamental domains, straightens out the physical interpretation and the topological origin of the corresponding invariants without resorting to more abstract K-theory arguments, and paves the way towards a deeper and more physically direct mathematical understanding of the impact of symmetries in topological band theory.

The remainder of this work is organised as follows: we begin in Sec.~\ref{sec:background} by providing the necessary theoretical background on Weyl semimetal topology, comparing the conventional physical approach with the (co)homology approach. We also offer a brief overview of the way in which non-orientable Brillouin zones arise, especially as it pertains to Weyl semimetals.
In Sec.~\ref{sec:motivation} we comment on some fundamental ambiguities in the current description of non-orientable Weyl semimetals, serving as a motivation for extending the coordinate-free (co)homology description into the non-orientable regime.
We then perform this extension by arguing how the (co)homology descriptions should be altered under orientation-reversing symmetries.
We apply these findings in Sec.~\ref{sec:k2s1} to provide a concrete classification of three-dimensional Weyl semimetals subject to a momentum-space glide symmetry.
Section~\ref{sec:extensions} revolves around several additional applications of this classification scheme, focussing on additional non-orientable Brillouin zones, non-Hermitian systems, and inversion-symmetric Weyl semimetals.
Our results are summarised and contextualised in Sec.~\ref{sec:sumdisout}, where we clarify important physical consequences and suggest avenues for future research.
We take stock and conclude in Sec.~\ref{sec:conclusion}.

\section{Background} \label{sec:background}
We devote this section to a brief background on two topics that are central to this work: (co)homology descriptions of the topological features of Weyl semimetals (Secs.~\ref{sec:WSMtopcohom}, \ref{sec:WSMtophom} and \ref{sec:bulkboundary}, based on Refs.~\cite{Mathai2017a,Mathai2017b}), and the emergence of non-orientable Brillouin zones (Sec.~\ref{sec:NOWSMs}, largely based on Refs.~\cite{Chen2022,fonseca2024}).
A more conventional discussion (from the point of view of a physicist) on Weyl semimetal topology is given in Sec.~\ref{sec:WSMtoppys}~\cite{weylreview}, allowing for a comparison between the physical and mathematical pictures.
A brief overview of relevant concepts in homology and cohomology is offered in Sec.~\ref{sec:hom-cohom}.
Readers who are familiar with these topics may proceed to Sec.~\ref{sec:motivation}, although it should be noted that the notation and other semantics introduced in this section will be used throughout the rest of the manuscript.

\subsection{Weyl semimetal topology: a physicist's approach} \label{sec:WSMtoppys}
A generic two-band system may be described by a Bloch Hamiltonian, the most general form of which takes the form
\begin{equation} \label{eq:BlochHamiltonian}
    \Ham(\bk) = d_0(\bk) \sigma^0 + \bd(\bk) \cdot \boldsymbol{\upsigma},
\end{equation}
where $\sigma^0$ denotes the $2\times 2$ identity matrix, $\boldsymbol{\upsigma} = (\sigma^x,\sigma^y,\sigma^z)$ is the vector of Pauli matrices, $\bk$ the lattice momentum, $d_0:\TT^d \to \RR$, and $\bd:\TT^d\to \RR^3$.
The energy bands of the system then correspond to the two eigenvalues of this Hamiltonian, which are given by
\begin{equation}
    E_{\pm}(\bk) = d_0(\bk) \pm \abs{\bd(\bk)} \coloneqq d_0(\bk) \pm \sqrt{\bd(\bk)\cdot\bd(\bk)}.
\end{equation}
The two bands touch when these eigenvalues are degenerate, which happens precisely when $\bd(\bk)=\mathbf{0}$.
In a $d$-dimensional system, this amounts to solving a system of three equations with $d$ parameters. For $d\geq 3$, such systems may have generic solutions that do not depend on external symmetries; these solutions result in so-called \emph{accidental band crossings}~\cite{weylreview}.
In three dimensions, these crossings tend to occur at isolated momenta called \emph{Weyl points} or \emph{Weyl nodes}, whose name stems from the fact that, to first order, the Hamiltonian at such a point resembles that of a chiral Weyl fermion.
These points are topological in nature, which means that they are stable under any sufficiently small perturbation, including (but not limited to) physical disorder.
The topology of a Weyl point is characterised by an integer invariant, related to the chirality of the corresponding Weyl fermion. This invariant can be computed as the flux of the Berry field through a sphere enclosing each point:
\begin{equation} \label{eq:chernnumber}
    \chi = \frac{1}{2\pi}\oint_{S^2} \Fc \in \ZZ,
\end{equation}
where $\Fc = \dd{\Ac}$ is the field strength of the Berry potential $\Ac = -i \braket{\psi}{\dd{\psi}}$ over the occupied bands.
Throughout the manuscript, this is what we will refer to as the notion of \emph{chirality}.
It should be noted that its sign depends on the orientation of the integration domain $S^2$: for example, mirroring the sphere reverses the sign of $\chi$.
The consequences of this will be highlighted later when studying Weyl nodes in non-orientable spaces.

In crystalline materials, the charges related to all Weyl nodes necessarily add up to zero across the Brillouin torus $\TT^3$, which can be seen from a Stokes' theorem argument;
\begin{equation} \label{eq:NNtheorem}
    \sum_{i=1}^k \chi_i = \frac{1}{2\pi}\sum_{i=1}^k\oint_{S^2_i}\Fc = \frac{1}{2\pi}\int_{\TT^3\setminus W}\dd{\Fc} = 0,
\end{equation}
where $W = \{w_1,...,w_k\}$ denotes the discrete set of Weyl points.
The last equality holds due to the Bianchi identity, which implies that the Berry field is divergence free.
This cancellation of charges is often referred to as the Nielsen--Ninomiya theorem~\cite{NM1981a,NM1981b,NM1983}.

\begin{figure}[htb!]
    \centering
	\def\svgwidth{.7\linewidth}
\begingroup%
  \makeatletter%
  \providecommand\color[2][]{%
    \errmessage{(Inkscape) Color is used for the text in Inkscape, but the package 'color.sty' is not loaded}%
    \renewcommand\color[2][]{}%
  }%
  \providecommand\transparent[1]{%
    \errmessage{(Inkscape) Transparency is used (non-zero) for the text in Inkscape, but the package 'transparent.sty' is not loaded}%
    \renewcommand\transparent[1]{}%
  }%
  \providecommand\rotatebox[2]{#2}%
  \newcommand*\fsize{\dimexpr\f@size pt\relax}%
  \newcommand*\lineheight[1]{\fontsize{\fsize}{#1\fsize}\selectfont}%
  \ifx\svgwidth\undefined%
    \setlength{\unitlength}{227.44653416bp}%
    \ifx\svgscale\undefined%
      \relax%
    \else%
      \setlength{\unitlength}{\unitlength * \real{\svgscale}}%
    \fi%
  \else%
    \setlength{\unitlength}{\svgwidth}%
  \fi%
  \global\let\svgwidth\undefined%
  \global\let\svgscale\undefined%
  \makeatother%
  \begin{picture}(1,0.54739157)%
    \lineheight{1}%
    \setlength\tabcolsep{0pt}%
    \put(0,0){\includegraphics[width=\unitlength,page=1]{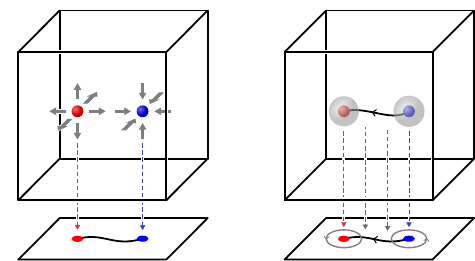}}%
    \put(-0.00425067,0.51350432){\color[rgb]{0,0,0}\makebox(0,0)[lt]{\lineheight{1.25}\smash{\begin{tabular}[t]{l}(a)\end{tabular}}}}%
    \put(0.55658027,0.51350432){\color[rgb]{0,0,0}\makebox(0,0)[lt]{\lineheight{1.25}\smash{\begin{tabular}[t]{l}(b)\end{tabular}}}}%
  \end{picture}%
\endgroup%

    \caption{A schematic comparison between (a) the conventional physical interpretation of Weyl point topology and (b) the corresponding (co)homology picture. In (a), a positively (negatively) charged Weyl point is depicted as a source (sink) of the Berry field inside the Brillouin zone. After projection to a surface, these nodes form the endpoints of a physical Fermi arc. In the cohomology picture (b), the charges correspond to second cohomology classes on spheres surrounding the Weyl points. 
    The oriented Dirac string connecting the two nodes represents a first homology counterpart to this: by Poincar\'e duality, the signed intersections of this string with the spheres agree with the cohomological charges.
    The Dirac string projects naturally onto a first homology class on the surface, which manifests as an oriented Fermi arc. Poincar\'e duality then ensures that the surface projections of the Weyl nodes have charges in terms of first cohomology classes on circles surrounding them.}
    \label{fig:physics_vs_cohomology}
\end{figure}

As is the case in topologically insulating materials, a bulk--boundary correspondence exists for Weyl semimetals: their bulk topology is reflected onto their surface, in the form of so-called \emph{Fermi arcs}. 
These dispersionless surface states define curves on the two-dimensional surface Brillouin zones in momentum space, bounded by the surface projection of the Weyl points; see Fig.~\ref{fig:physics_vs_cohomology}(a).
These can be thought of as projections of the Fermi energy of two-dimensional gapped planes perpendicular to the surface, located between the Weyl points.

\subsection{Homology, cohomology and their extensions} \label{sec:hom-cohom}

Here we briefly review the concepts of homology and cohomology, along with some extensions used in this work.
This section is meant to be a conceptual overview; for a technical introduction, the reader is referred to standard textbooks such as Refs.~\cite{Hatcher,BottTu}.

\subsubsection{Homology} \label{sec:homology}
In classifying the topology of a space $M$, it can be helpful to consider which of its subspaces are ``non-trivial'' in the sense that they cannot be continuously contracted to a point or deformed into one another. Conceptually, perhaps the simplest way to approach this is using the closely related \emph{homotopy} theory: the $n$-th homotopy group $\pi_n(M)$ contains classes of topologically distinct maps from the $n$-dimensional sphere $S^n$ into $M$. In particular, $\pi_1(M)$ is called the fundamental group of $M$ and essentially classifies winding numbers of 1-dimensional loops around different internal structures in $M$. For example, in the three-dimensional Brillouin torus $\TT^3$, a loop spanning the space in the $k_x$-direction represents a different homotopy class from one that encircles $\TT^3$ twice in the same direction, or one running in a different coordinate direction.

Despite their conceptual simplicity, homotopy groups can be troublesome to compute and interpret even for relatively simple spaces, due to the difficulty of keeping track of all the necessary maps and their deformations. This is less of a problem in \emph{homology} theory, where the non-trivial subspaces are defined using the concept of boundaries: in basic terms, a non-zero homology class corresponds to a subspace which has no boundary (e.g.\ a closed loop), but which is itself not the boundary of some higher-dimensional subspace (e.g.\ a two-dimensional disc has a loop as a boundary, but this can be contracted to a point along the disc). A similar concept to the winding numbers of homotopy is then instead recovered in terms of \emph{coefficients} from some commutative group $G$, most commonly integers. For example, an element of the first homology group $H_1(\TT^3;G)$ is represented (up to deformation) by a sum of loops of the form
\begin{equation} \label{eq:coefficients}
    a\ell_x + b\ell_y + c\ell_z, \quad \{a,b,c\}\in G,
\end{equation}
where $\ell_i$ spans the $k_i$-direction once. It turns out that choosing integer coefficients $G=\ZZ$ provides the richest topological information, and it is common to abbreviate $H_n(M):=H_n(M;\ZZ)$. The coefficients give the homology group its group structure: in the case of $\TT^3$, one finds that $H_1(\TT^3)\cong\ZZ^3$, where each $\ZZ$ factor is generated by one of the three basic loops $\ell_i$. The elements of this group are called \emph{first homology classes}, and we will denote them by bracketed representatives, e.g.\ $[\ell_x + 2\ell_y] = [\ell_x] + 2[\ell_y] \in H_1(\TT^3)$.

Homology classes may have both positive and negative coefficients, and this instils a sense of orientation: for example, $-\ell_x$ may be thought of as a loop spanning $\TT^3$ in the negative $k_x$-direction. This orientation must be respected in a particular by the boundary map $\partial$ which is used to define the homology groups: most importantly for this work, a one-dimensional string $l$ which runs away from a point $p$ towards another point $q$ has the formal boundary $\partial l = q - p$.

One particular refinement of note in this work is the notion of homology \emph{relative} to a subspace $A\subset M$: the relative homology groups $H_n(M,A)$ classify subspaces which are allowed to have boundaries in $A$. For example, if $A$ consists of the two points $p$ and $q$ mentioned above, then the string $l$ running between them represents an element of $H_1(M,A)$.

\subsubsection{Cohomology} \label{sec:cohomology}

Beyond computability, a key advantage of homology is that it can be dualised into a physically natural counterpart called cohomology. This dualisation is performed in terms of linear maps which map the homological (sums of) oriented subspaces to the coefficient group $G$; the duality arises because such a map can be paired with a single subspace to give a specific element of $G$. The most physically familiar example of this comes in the form of \emph{de Rham cohomology} for the real coefficients $G=\RR$. In this instance, the linear maps under consideration are differential forms; for example, the Berry field $\Fc$ can be considered a curvature 2-form. The dual pairing then takes the form of integration; for example, in Eq.~\eqref{eq:chernnumber}, the Berry field pairs with a two-sphere to return a real multiple of $2\pi$.

Similarly to homology, these structures are ordered into cohomology groups $H^n$ using a \emph{coboundary} map. In the example of de Rham cohomology, this is the exterior derivative $\dd$, and the $n$-th cohomology group $H^n(M;\RR)$ classifies ways in which closed $n$-forms $\omega$ (i.e.\ those obeying $\dd\omega = 0$) may fail to be globally exact (i.e.\ of the form $\omega = \dd\eta$ for some $(n-1)$-form $\eta$). The Berry curvature 2-form $\Fc$, for example, may always be written as $\Fc=\dd\Ac$ \emph{locally}, but the connection 1-form $\Ac$ may be subject to gauge singularities \emph{globally}, requiring multiple different gauge choices on overlapping domains. In this case the system is topologically non-trivial.

As with homology, the real coefficients of de Rham cohomology may not always capture the full topology of a system, and we will implicitly make use of integer coefficients in this work.
In this setting, the second cohomology group $H^2(M)$ is often of special interest: second cohomology classes precisely classify complex line bundles over the base space $M$. These may be interpreted as so-called \emph{valence bundles} of topologically distinct insulating two-band systems, offering a different (but equivalent) approach to topology from the physical notion of adiabatic connectedness.

Aside from the defining duality between homology and cohomology, there is another so-called \emph{Poincar\'e duality} which is important in this context: if $M$ is a $d$-dimensional closed \emph{oriented} manifold, then there is an isomorphism
\begin{equation} \label{eq:Poincareduality}
    H_n\left(M\right) \cong H^{d-n}\left(M\right)
\end{equation}
for any integer $n$. If the the homology group on the left hand side is instead taken relative to some subset $A\subset M$, then the cohomology group on the right must be modified to exclude $A$ accordingly: the statement then becomes
\begin{equation} \label{eq:Lefschetzduality}
    H_n\left(M,A\right) \cong H^{d-n}\left(M\setminus A\right),
\end{equation}
sometimes known as \emph{Lefschetz duality}.

\subsubsection{Equivariant and twisted (co)homology} \label{sec:homology-modifications}

There are two modifications to regular homology and cohomology that play a role in this work.

The first of these is \emph{equivariant (co)homology}, which is used to describe systems subject to symmetries.
Given a symmetry group $G$ acting on a space $M$, the $n$-th $G$-equivariant homology group $H_n^G(M)$ (upper and lower index reversed for cohomology) can be thought of as classifying oriented $n$-dimensional topological subspaces of $M$ which are fixed by the symmetry.
In cases where the symmetry has individual fixed points, this requires the use of a special construction called the \emph{homotopy quotient}, which encodes all the necessary topological information.
However, most symmetries under consideration in this work have no fixed points, and in these cases one can equivalently work with regular (co)homology on the quotient space $M/G$ of the symmetry~\cite{Tu2020}:
\begin{equation} \label{eq:equivariant-homology}
    H_n^G(M) \cong H_n(M/G),
\end{equation}
and similarly for cohomology.
In our case, this quotient space takes the form of a reduced Brillouin zone.

The second modification, which is central to this work, is \emph{twisted (co)homology}, which can be used to counter the effects of non-orientability -- in particular, to remedy the lack of Poincar\'e duality in this context.
The idea is that, instead of the fixed coefficient group appearing in Eq.~\eqref{eq:coefficients}, so-called \emph{local coefficients} can be used that are ``attached'' to the base space $M$ in a way which may depend non-trivially on the global topology of $M$.
This is achieved by letting the fundamental group $\pi_1(M)$ (introduced in Sec.~\ref{sec:homology} above) act on the group of coefficients in some way.
We use a common choice of such local coefficients in this work: we notate $\widetilde{\ZZ}:= \ZZ^\omega$ for integer coefficients with a fundamental group action defined by the \emph{orientation character} $\omega:\pi_1(M)\to\{\pm 1\}$, where orientation-reversing loops are mapped to -1. Effectively, this ensures that moving along an orientation-reversing path induces a sign change on the (co)homology elements.

This specific choice of local coefficients gives rise to a more general \emph{twisted Poincar\'e duality} which does not depend on orientability. To be precise, the following two statements hold on any closed $d$-dimensional manifold, regardless of orientability~\cite{Hatcher}:
\begin{equation} \label{eq:twisted-poincare}
    H_n\left(M;\ZZtwist\right)\cong H^{d-n}\left(M\right), \quad H^n\left(M;\ZZtwist\right)\cong H_{d-n}\left(M\right).
\end{equation}
That is, only one of the groups must be twisted to restore duality: twisted homology is Poincar\'e dual to ordinary cohomology and vice versa. In the case when $M$ is an oriented manifold, the local coefficients $\widetilde{\ZZ}$ are equivalent to normal integer coefficients, and both statements reduce to the ordinary Poincar\'e duality of Eq.~\eqref{eq:Poincareduality}.

\subsection{Weyl semimetal topology: a cohomology approach} \label{sec:WSMtopcohom}
We now review an approach to Weyl semimetal topology using the (co)homology discussed in Sec.~\ref{sec:hom-cohom}, based on Ref.~\cite{Mathai2017a,Mathai2017b}. 
Note that the description provided in the next few sections assumes that no additional symmetries beyond lattice translation are enforced on the system; going beyond this requires the modifications discussed in Sec.~\ref{sec:homology-modifications}, and we work this out for our purposes in Sec.~\ref{sec:NOWSMs}.

We begin from the cohomology point of view, which is closest to the physical picture. 
We will review the specifics here in terms of the familiar language of de Rham cohomology and Berry curvatures, which is adequate in the low-symmetry case. 
However, the reader should keep in mind that integer coefficients are used implicitly, and the intuition of Berry curvature is incomplete when e.g.\ $\ZZ_2$ invariants become involved, which cannot be expressed as simple integrals on surfaces.

As discussed in Sec.~\ref{sec:cohomology}, topological systems can be classified in terms of second cohomology groups.
Since the Berry curvature has a singularity at each Weyl point, the corresponding topology is classified by curvatures away from these points; as a result, the topology of Weyl semimetals is classified by second cohomology on the punctured space $\TT^3\setminus{W}$, i.e.\ the Brillouin zone $\TT^3$ with the collection of Weyl points $W$ removed.
In particular, a three-dimensional Chern insulator can be considered a special case of a Weyl semimetal with zero Weyl points; as a result, its topology is classified by the second cohomology group of the full (unpunctured) torus $\TT^3$.

The physical information contained in these groups is extracted using a \emph{Mayer--Vietoris sequence}, a widely used tool within algebraic topology relating the cohomology of a space to that of a choice of subspaces~\cite{Hatcher}.
By making the appropriate choice of subspaces, these sequences relate the topological charge of each individual Weyl point to the global Weyl semimetal topology~\cite{Mathai2017a,Mathai2017b}.
The common choice of subspaces covers the 3-torus as follows:
\begin{equation}
    \TT^3 = \left(\TT^3\setminus W\right) \cup \coprod_{i=1}^k D^3_i,
\end{equation}
i.e.\ the Brillouin torus $\TT^3$ is covered by the punctured torus $\TT^3\setminus W$ with the Weyl points removed, and a disjoint union of solid balls (3-discs) $D^3_i$ centred on the respective Weyl points.
The relevant part of the corresponding Mayer--Vietoris sequence then reads
\begin{equation}\label{eq:semimetal-MV-symbolic}  
	0\ \to\ \underbrace{H^2\left(\TT^3\right)}_{\mathclap{\text{3D Chern insulator}}}\ \to\ 
	\underbrace{H^2\left(\TT^3\setminus W\right)}_{\text{Semimetal}}\ \overset{\beta}{\to}\ \underbrace{\bigoplus_{w_i\in W} H^2\left(S_{w_i}^2\right)}_{\text{Local charges}}\ \overset{\Sigma}{\to}\ \underbrace{H^3\left(\TT^3\right)}_{\mathclap{\text{Total charge}}}\ \to\ 0.
\end{equation}
Before moving on, let us comment on the physical meaning of each group individually. 
Starting on the left, $H^2\left(\TT^3\right)$ is exactly what provides the topology of a three-dimensional Chern insulator, which is given by a vector of three distinct Chern numbers $\vb{C} = (C_x,C_y,C_z)$~\cite{Devescovi}.
These Chern numbers are calculated as Berry curvature integrals on two-dimensional toroidal slices of the Brillouin zone, similar to the Weyl point charge in Eq.~\eqref{eq:chernnumber}: \begin{equation} \label{eq:insulatorchern}
    C_i = \frac{1}{2\pi}\oint_{\TT_{k_i=\text{const.}}^2} \Fc \in \ZZ,
\end{equation}
where the integration domain is any slice of constant $k_i$.
Second, the group $H^2\left(\TT^3\setminus W\right)$ classifies all possible semimetallic phases given a fixed set of Weyl points $W$.
Third, the collection of $H^2\left(S^2_{w_i}\right)$ gives information about the local charges of each Weyl point $w_i\in W$.
The map $\beta$ into this group can then be thought of as assigning a list of Weyl point charges to each specific semimetal configuration.
Finally, the third cohomology group $H^3\left(\TT^3\right)$ can be thought of as classifying total charge on the Brillouin torus, with the map $\Sigma$ acting as a sum over the individual charges; as a result, non-zero elements of this group represent unphysical configurations.

The groups in Eq.~\eqref{eq:semimetal-MV-symbolic} can be calculated explicitly, after which the sequence takes the form
\begin{equation}\label{eq:semimetal-MV-explicit}
	0 \to \ZZ^3 \to \ZZ^3\oplus\ZZ^{k-1} \overset{\beta}{\to} \ZZ^k \overset{\Sigma}{\to} \ZZ \to 0,
\end{equation}
where $k=\abs{W}$ is the number of Weyl points in the system.
The relation between local and global topological properties is provided by the maps between these groups.
These maps are not independent of each other; Mayer--Vietoris sequences are known to be \emph{exact}, meaning that the image of one map is always equal to the kernel of the next map in the sequence.
This exactness sets restrictions on the way in which the elements of each group are mapped into the subsequent one.

There are two features of this sequence that are of particular interest from the perspective of semimetal topology. The first of these relates to the maps $\beta$ and $\Sigma$ around the group of local charges $\ZZ^k$ in Eq.~\eqref{eq:semimetal-MV-explicit}; here, exactness implies that $\im(\beta) = \ker(\Sigma)$. That is, the local Chern numbers sum to zero if and only if they descend from a semimetal structure.
This tells us two different things: the first is that the sum of Weyl point chiralities in a Weyl semimetal is necessarily zero; this is precisely the content of the Nielsen--Ninomiya theorem. 
In addition, it also implies the converse: any configuration of charges that sums to zero is necessarily compatible with global semimetal topology, i.e.\ there are no such configurations which do not appear in any possible semimetal.

The second important feature of this sequence lies in the nature of the semimetal group $H^2\left(\TT^3\setminus W\right)$ itself.
Its appearance between a group of insulating invariants and the charge cancellation construction discussed above signifies that semimetal topology can be fundamentally understood as a product of these two aspects.
This is precisely the reason for the suggestive notation $\ZZ^3\oplus\ZZ^{k-1}$ of this group: for any given set of $k$ Weyl points there are $\ZZ^{k-1}$ valid charge configurations (i.e.\ $k$ different $\ZZ$-valued charges subject to one $0\in\ZZ$ charge cancellation constraint), and for any such configuration, the topology is further determined by a $\ZZ^3$ background of insulating invariants of the form of the Chern numbers $C_i$ of Eq.~\eqref{eq:insulatorchern}.

\subsection{Weyl semimetal topology: a homology approach} \label{sec:WSMtophom}
Given the cohomology approach outlined above, one can obtain a corresponding \emph{homology} picture by applying the Poincar\'e duality of Eqs.~\eqref{eq:Poincareduality} and \eqref{eq:Lefschetzduality}.
This duality allows the second cohomology groups involved in classifying invariants to be reframed as first homology groups: that is, there is a classification based on oriented one-dimensional subspaces of the Brillouin zone.
The semimetal Mayer--Vietoris sequence in Eq.~\eqref{eq:semimetal-MV-symbolic} dualises to the following form:
\begin{equation}\label{eq:semimetal-homology-sequence}
	0\ \to\ \underbrace{H_1\left(\TT^3\right)}_{\mathclap{\text{Dirac loops}}}\ \to\ 
	\underbrace{H_1\left(\TT^3, W\right)}_{\text{Dirac strings}}\ \overset{\partial}{\to}\ \underbrace{H_0\left(W\right)}_{\mathclap{\text{Local charges}}}\ \overset{\Sigma}{\to}\ \underbrace{H_0\left(\TT^3\right)}_{\mathclap{\text{Total charge}}}\ \to\ 0.
\end{equation}
Note that this is a valid exact sequence in homology independently of the validity of Poincar\'e duality in a given setting; the duality only ensures that it classifies the same topology as the cohomology sequence.

Special attention should be paid to the second group in this sequence, which is obtained by the Lefschetz duality of Eq.~\eqref{eq:Lefschetzduality} from the semimetal invariant group $H^2\left(\TT^3\setminus W\right)$. 
The resulting relative homology group $H_1\left(\TT^3,W\right)$ classifies oriented one-dimensional spaces that are allowed to terminate at the Weyl points; an example of this is shown in Fig.~\ref{fig:physics_vs_cohomology}(b).
Such spaces are called \emph{Dirac strings}, in reference to the fact that they can be physically considered to be strings of gauge singularities running between Weyl nodes.

The map $\partial$ appearing in Eq.~\eqref{eq:semimetal-homology-sequence} is the defining boundary map in homology. As discussed in Sec.~\ref{sec:homology}, this map always assigns opposite signs to the endpoints of a string, ensuring that Dirac strings always connect two oppositely charged Weyl points.
As a result, the Nielsen--Ninomiya theorem is encoded in the homology sequence in a very natural way: the opposite charges bounding a Dirac string sum to zero by nature when the map $\Sigma$ is applied.

The first group in the sequence classifies the insulating topology in terms of what we will refer to as \emph{Dirac loops}.
These can be thought of as Dirac strings which are closed into loops along the periodic directions of the torus, and indeed such loops may form when Weyl nodes are annihilated together along topologically non-trivial paths.
In this regard, Weyl semimetals can be considered transitional phases between different insulating topological phases; this is precisely why a group of insulating invariants appears in each of the exact sequences under consideration.

Finally, the Poincar\'e duality that relates the homology and cohomology classifications has a natural geometric consequence in any given topological phase: for any two-dimensional surface within the Brillouin zone, the (cohomological) Chern number on that surface agrees with the number of signed intersections it has with (homological) oriented Dirac strings or loops.
For example, a Dirac loop that spans $\TT^3$ once in the $k_x$-direction precisely induces a unit Chern number $C_x$, as defined in Eq.~\eqref{eq:insulatorchern}.
Similarly, the orientation of a Dirac string always agrees with the cohomological charge of the Weyl nodes at either end, defined on the 2-spheres surrounding them -- this is also illustrated in Fig.~\ref{fig:physics_vs_cohomology}(b).
Seeing as Poincar\'e duality depends on orientability of the Brillouin zone, this is one of the aspects that needs to be treated carefully in the non-orientable case.

\subsection{Bulk--boundary correspondence via homology} \label{sec:bulkboundary}
As noted in Sec.~\ref{sec:WSMtoppys}, the bulk topology of Weyl semimetals gives rise to surface Fermi arcs via the bulk--boundary correspondence. This correspondence can be made precise using the language of homology.
To this end, assume the existence of a projection $\pi:\TT^3\to\TT^2$ that maps the bulk onto a two-dimensional surface Brillouin torus, and let $W' \coloneqq \pi(W)$ denote the surface projection of the set of Weyl nodes $W$. 
The projection $\pi$ induces a map $\pi_*$ between homology groups, called the \emph{pushforward} on homology classes:
\begin{equation}
    \pi_*: H_1\left(\TT^3,W\right)\to H_1\left(\TT^2,W'\right),\quad [s]\mapsto\left[\pi(s)\right],
\end{equation}
where $[s]$ denotes the homology class represented by a Dirac string configuration $s$.
As it turns out, the resulting surface homology class $\pi_*\left(\left[s\right]\right) \coloneqq \left[\pi\left(s\right)\right]$ is precisely the homology class of the physical Fermi arc~\cite{Mathai2017a,Mathai2017b,Gomi2020}. In other words, loosely speaking, Fermi arcs can be considered surface projections of bulk Dirac strings; this is illustrated in Fig.~\ref{fig:physics_vs_cohomology}(b).

One can apply similar reasoning to the rest of the exact sequence in Eq.~\eqref{eq:semimetal-homology-sequence}, resulting in a similar exact sequence on the surface:
\begin{equation} \label{eq:surface-homology-sequence}
    0\ \to\ \underbrace{H_1\left(\TT^2\right)}_{\mathclap{\text{Fermi loops}}}\ \to\ 
	\underbrace{H_1\left(\TT^2, W'\right)}_{\text{Fermi arcs}}\ \overset{\partial}{\to}\ \underbrace{H_0\left(W'\right)}_{\mathclap{\text{Surface charges}}}\ \overset{\Sigma}{\to}\ H_0\left(\TT^2\right)\ \to\ 0.
\end{equation}
Calculated in terms of explicit groups, this sequence takes on a similar form to Eq.~\eqref{eq:semimetal-MV-explicit}:
\begin{equation}\label{eq:surface-MV-explicit}
	0 \to \ZZ^2 \to \ZZ^2\oplus\ZZ^{k'-1} \overset{\partial}{\to} \ZZ^{k'} \overset{\Sigma}{\to} \ZZ \to 0,
\end{equation}
where $k' = |W'|$ is the number of projected Weyl points; note that this may be less than the number of bulk points in fine-tuned scenarios where the Weyl points overlap along the exact direction of $\pi$. 
In these cases the surface topology behaves exactly as though there were fewer bulk Weyl points.

This sequence is in many ways analogous to its bulk counterpart; in particular, the boundary map $\partial$ and the summation $\Sigma$ retain the same interpretation, indicating that the Nielsen--Ninomiya theorem is also topologically enforced on the surface.
The \emph{Fermi loops} appearing in the leftmost group are boundary states of three-dimensional Chern insulators, which result from closing Fermi arcs into non-trivial loops -- a process which has been observed experimentally~\cite{Liu2022}.

Finally, the surface topology can also be framed in terms of cohomology, allowing for a more direct physical interpretation.
This is achieved by another application of Poincaré duality, with the key difference that the first homology groups are dualised to \emph{first} instead of second cohomology groups due to the two-dimensional nature of the surface Brillouin zone.
The dual of the homology sequence in Eq.~\eqref{eq:surface-homology-sequence} is then equivalent to the following Mayer--Vietoris sequence:
\begin{equation}
    \dots\ \overset{0}{\to}\ \underbrace{H^1\left(\TT^2\right)\ \to\ 
	H^1\left(\TT^2\setminus W'\right)}_{\text{Gapless boundary states}}\ \to \underbrace{\bigoplus_{\tilde{w}_i\in W'} H^1\left(S_{\tilde{w}_i}^1\right)}_{\text{Surface charges}}\ \overset{\Sigma}{\to}\ H^2\left(\TT^2\right)\ \to\ 0,
\end{equation}
where the two groups on the left hand side precisely classify the different surface states of the three-dimensional Chern insulators and semimetals appearing in Eq.~\eqref{eq:semimetal-MV-symbolic}.
As a small note, the group to the left of this sequence is formally non-zero in the full Mayer--Vietoris sequence due to the low dimensionality, but since it maps trivially into $H^1(\TT^2)$, the interpretation is unchanged.

Physically speaking, the appearance of first cohomology groups in this sequence implies that the surface invariants in these systems can be calculated by integration of 1-forms; for example, as illustrated in Fig.~\ref{fig:physics_vs_cohomology}(b), the topological charge of a projected Weyl node $\tilde{w}$ can be calculated as a winding number of the Berry phase on a small circle $S_{\tilde{w}}^1$ enclosing it~\cite{weylreview}.
We should be careful to note that this integration is not performed on the boundary of a Hamiltonian with open boundary conditions, where the winding number would be ill-defined because of the existence of the gapless Fermi arcs.
Rather, the winding number is calculated using the two-dimensional Bloch Hamiltonian obtained by projecting the three-dimensional bulk Hamiltonian along the map $\pi$.
In this projected Hamiltonian, the gapless Fermi arcs obtained using open boundary conditions correspond to strings of gauge singularities, analogous to Dirac strings in the bulk, which are classified using first homology~\cite{Mathai2017b}.

\subsection{Non-orientable systems} \label{sec:NOWSMs}
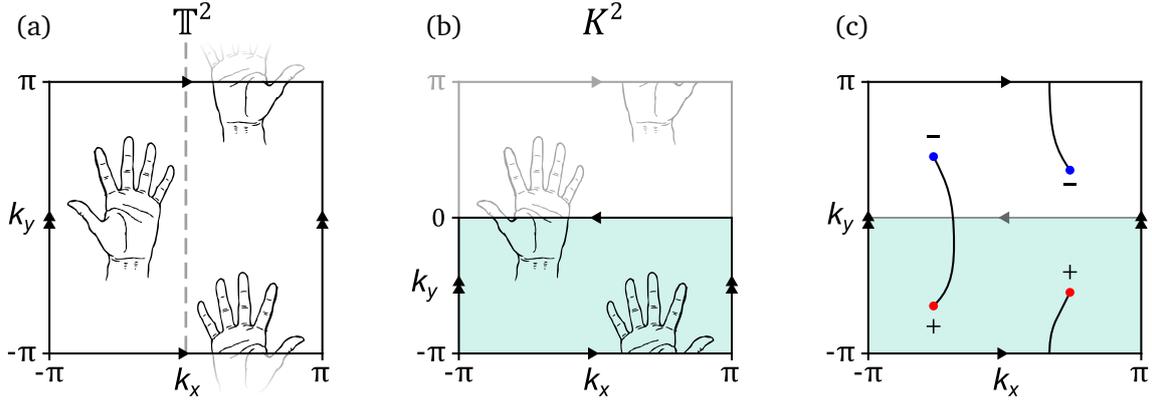
\begin{figure}
    \centering
	\def\svgwidth{\linewidth}
\begingroup%
  \makeatletter%
  \providecommand\color[2][]{%
    \errmessage{(Inkscape) Color is used for the text in Inkscape, but the package 'color.sty' is not loaded}%
    \renewcommand\color[2][]{}%
  }%
  \providecommand\transparent[1]{%
    \errmessage{(Inkscape) Transparency is used (non-zero) for the text in Inkscape, but the package 'transparent.sty' is not loaded}%
    \renewcommand\transparent[1]{}%
  }%
  \providecommand\rotatebox[2]{#2}%
  \newcommand*\fsize{\dimexpr\f@size pt\relax}%
  \newcommand*\lineheight[1]{\fontsize{\fsize}{#1\fsize}\selectfont}%
  \ifx\svgwidth\undefined%
    \setlength{\unitlength}{473.58730046bp}%
    \ifx\svgscale\undefined%
      \relax%
    \else%
      \setlength{\unitlength}{\unitlength * \real{\svgscale}}%
    \fi%
  \else%
    \setlength{\unitlength}{\svgwidth}%
  \fi%
  \global\let\svgwidth\undefined%
  \global\let\svgscale\undefined%
  \makeatother%
  \begin{picture}(1,0.34720794)%
    \lineheight{1}%
    \setlength\tabcolsep{0pt}%
    \put(0.72466803,0.31888608){\color[rgb]{0,0,0}\makebox(0,0)[lt]{\lineheight{1.25}\smash{\begin{tabular}[t]{l}(c)\end{tabular}}}}%
    \put(0.00641085,0.31888608){\color[rgb]{0,0,0}\makebox(0,0)[lt]{\lineheight{1.25}\smash{\begin{tabular}[t]{l}(a)\end{tabular}}}}%
    \put(0.36553944,0.31888608){\color[rgb]{0,0,0}\makebox(0,0)[lt]{\lineheight{1.25}\smash{\begin{tabular}[t]{l}(b)\end{tabular}}}}%
    \put(0,0){\includegraphics[width=\unitlength,page=1]{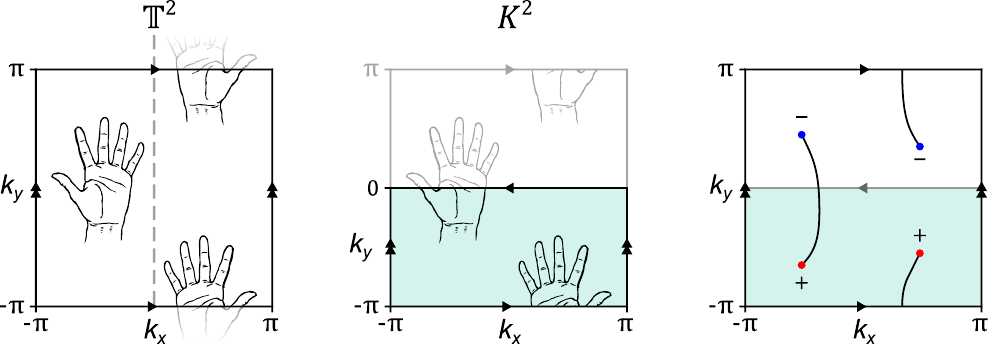}}%
  \end{picture}%
\endgroup%

    \caption{Illustration of the effect of glide symmetry on a two-dimensional Brillouin zone.
    (a) The symmetry acts on objects in the Brillouin zone by simultaneous reflection and translation, changing their handedness. On the full torus $\TT^2$, the chiralities of the resulting objects are still well defined.
    (b) After reduction to the shaded fundamental domain, the Brillouin zone becomes non-orientable. The notion of handedness is lost in the process: the left palm traversing the orientation-reversing line at $k_y=0$ emerges as a right palm at $k_y=-\pi$. 
    (c) Equivalent behaviour appears on the surface Brillouin zone of a glide symmetric Weyl semimetal: from the perspective of the shaded fundamental domain, a Fermi arc which crosses $k_y=0$ appears to connect similarly charged Weyl nodes.}
    \label{fig:BZ_glide_symmetry}
\end{figure}

As noted earlier, exposing a system to various symmetries may drastically alter its topological properties.
In this work, we are concerned with orientation-reversing symmetries, and in particular those which give rise to alternate Brillouin zone topologies.
Such alternate Brillouin zones were first explored in two dimensions in Ref.~\cite{Chen2022}, in which gauge fluxes on a real-space lattice give rise to a so-called \emph{glide symmetry} in momentum space. This symmetry combines a reflection along the $k_x$-axis with a half-lattice translation along the $k_y$-axis; see Fig.~\ref{fig:BZ_glide_symmetry}(a). Explicitly, the momentum-space Hamiltonian is subject to the constraint
\begin{equation} \label{eq:2DGlide}
    \Ham\left(k_x,k_y\right) = U\Ham\left(-k_x,k_y+\pi\right)U^{-1}.
\end{equation}
A special property of this glide symmetry is that its action on the Brillouin torus $\TT^2$ is free, i.e.\ it leaves no points fixed. As a result, the physics of the system is completely captured on a \emph{fundamental domain} covering precisely half of the torus with $-\pi \leq k_y \leq 0$, which thus acts as the new Brillouin zone. Under the correct boundary identifications, this fundamental domain takes on the topology of the non-orientable Klein bottle $K^2$; see Fig.~\ref{fig:BZ_glide_symmetry}(b). In this specific instance, the loss of orientability changes the $\ZZ$-valued Chern number that would normally be expected in the system to an invariant valued in $\ZZ_2=\{0,1\}$; that is, the system has only one trivial and one topological phase.

It should be noted that the appearance of this glide symmetry is exceptional, in the sense that real-space symmetries usually give rise to \emph{symmorphic} symmetries which do not include half-lattice translations. 
However, the lattice gauge fluxes of Ref.~\cite{Chen2022} provide a plausible physical basis, and it has since been shown how any non-symmorphic momentum-space symmetry can be induced similarly in two and three dimensions~\cite{Zhang2023}. 
While still elusive in electronic materials, such momentum-space symmetries have been experimentally realised with great success in artificial materials such as acoustic crystals~\cite{Tao2024,Zhu2024,Wang2023}.

One such three-dimensional symmetry has recently been applied to the setting of Weyl semimetals in Ref.~\cite{fonseca2024}, where a three-dimensional glide symmetry is imposed in the form of the following constraint:
\begin{equation} \label{eq:3DGlide}
    \Ham\left(k_x,k_y,k_z\right) = U\Ham\left(-k_x,k_y+\pi,k_z\right)U^{-1}.
\end{equation}
This glide symmetry acts analogously to the one above, yielding a Klein bottle topology in the first two coordinates; with the addition of the periodic $k_z$ direction, the final topology of the Brillouin zone becomes $\KS$.

The loss of orientability of the Brillouin zone causes the chiral nature of Weyl nodes to become less well-defined, leading to novel topological features.
For example, moving a Weyl node with Chern number $+1$ out of the fundamental domain across the $k_y = -\pi$ plane will cause it to reappear as a Weyl point with Chern number $-1$ at $k_y = 0$.
Consequently, the sum of all the Chern numbers associated to the Weyl nodes in the fundamental domain may no longer add to zero, seemingly circumventing the conventional Nielsen--Ninomiya theorem.
Instead, Weyl points emerging in a $\KS$ fundamental domain are found to satisfy an altered $\ZZ_2$ cancellation theorem, meaning that over a fundamental domain $\KS$, the charges of a set of Weyl points $W$ are constrained to satisfy
\begin{equation}
    \sum_{w_i\in W} \chi_i = 0 \mod 2.
\end{equation}
This has a further natural manifestation in terms of the Fermi arcs on the $K^2$-like $\left(k_x,k_y\right)$-surfaces: instead of only connecting Weyl points of opposite chiralities, Fermi arcs may be bounded by same-chirality Weyl points. Specifically, this is found to occur when a Fermi arc crosses the orientation-reversing $k_y=0$ boundary of the surface Brillouin zone an odd number of times; see Fig.~\ref{fig:BZ_glide_symmetry}(c).
In the next section, we argue that the physical interpretation of these features is confounded by fixing a specific fundamental domain, and instead develop a coordinate-free formalism based on the (co)homology approach described above.

\section{Non-orientable (co)homology formalism} \label{sec:motivation}

\begin{figure}[htb!]
	\centering
	\subcaptionbox{$-\pi \leq k_y \leq 0$\label{subfig:BZ_basic}} {\includegraphics[width=.3\textwidth]{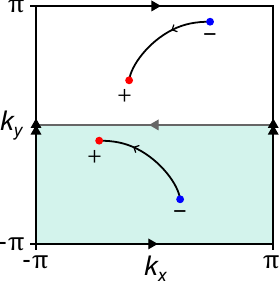}}
	\hfil
	\subcaptionbox{$-\pi/2 \leq k_y \leq \pi/2$\label{subfig:BZ_mid}}{\includegraphics[width=.3\textwidth]{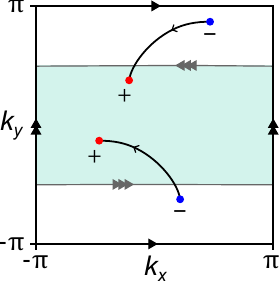}}
	\hfil
	\subcaptionbox{$0 \leq k_x \leq \pi$\label{subfig:BZ_right}} {\includegraphics[width=.3\textwidth]{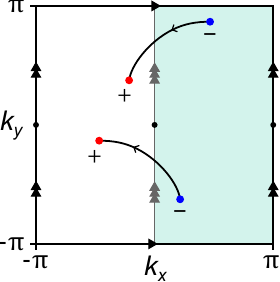}}
	\caption{Top view of the three-dimensional Brillouin torus (or two-dimensional surface torus) for a given Weyl semimetal subject to the glide symmetry in Eq.~\eqref{eq:3DGlide}. A simple configuration of Weyl nodes is shown, with oriented Dirac strings (Fermi arcs) connecting them. Different choices for the fundamental domain are shaded in teal, all of which take on $\KS$ ($K^2$) topology after identifying their boundaries along the indicated arrows: (a) the domain outlined in Ref.~\cite{fonseca2024}; (b) the same domain shifted in the $k_y$ direction; (c) a domain spanning the $k_y$ direction. Consequently, the notions of both relative and absolute chirality of the two Weyl nodes in the fundamental domain are affected by its position within the torus, indicating that these notions do not have phenomenological implications.}
	\label{fig:BZ_param}
\end{figure}

The glide symmetry defined in Eq.~\eqref{eq:3DGlide} leads to fascinating topological behaviour, but the discussion of this behaviour is somewhat hampered in the context of a fixed fundamental domain. For example, it is unclear in this context whether the possibly non-zero total charge appearing in the fundamental domain has any implications for the physical properties of the corresponding semimetals. However, as shown in Fig.~\ref{fig:BZ_param}, the total Weyl point charge may depend sensitively on a given choice of the fundamental domain within the Brillouin torus.

To alleviate the potential confusion arising from this choice of fundamental domain, we strive to give an appropriate description of these non-orientable Weyl semimetals in terms of the more general (co)homology framework in this section, which forms the basis of our results. We will initially treat free orientation-reversing symmetries in full generality, without assuming the specific glide symmetry or the unitarity implicit in Eq.~\eqref{eq:3DGlide}.

Before we move on, we include a small note on the Poincar\'e--Hopf theorem, which is often used to derive the Nielsen--Ninomiya theorem.
In the usual argument, the map $\bd(\bk)$ which sets the Pauli matrix coefficients in the Bloch Hamiltonian [Eq.~\eqref{eq:BlochHamiltonian}] is treated as a vector field on $\TT^3$; Weyl nodes then correspond to singularities of this vector field.
The Poincar\'e--Hopf theorem then states that the indices of these singularities (i.e.\ the chiralities of the Weyl nodes) must add to zero.
This argument implicitly relies on the fact that $\TT^3$ can be parametrised at once by three (periodic) global coordinates, ensuring that the map $\bd: \TT^3\to\RR^3$ can be identified with a vector field directly.
As such, the argument does not generalise straightforwardly to other Brillouin zone topologies: in particular, non-orientable manifolds such as the $\KS$ topology studied in Ref.~\cite{fonseca2024} never admit global coordinates, and $\bd:\KS\to\RR^3$ is fundamentally different from a vector field on $\KS$.
In technical terms, a vector field on a $d$-dimensional manifold $M$ is a section of its \emph{tangent bundle}, whereas a map $M\to\RR^d$ is a section of the \emph{trivial bundle} $M\times\RR^d$. In order for the argument to work, these two bundles must be equivalent; in this case, $M$ is called \emph{parallelisable}.
In three dimensions, the situation is straightforward: a three-dimensional closed manifold is parallelisable if and only if it is orientable. This means the Poincar\'e--Hopf argument requires modification to be applied in the non-orientable case -- a point of view which we will not pursue further, in favour of the richer (co)homology interpretation.

The presence of symmetries will generally enforce modifications of the (co)homology treatment reviewed in Secs.~\ref{sec:WSMtopcohom} and \ref{sec:WSMtophom}. These modifications come in the form of the equivariant and twisted groups introduced in Sec.~\ref{sec:homology-modifications}. The applications of such constructions to topological matter have been well studied in the more general and more abstract setting of K-theory~\cite{Freed2013,Thiang2015,Gomi2015,Shiozaki2022,Gomi2017}, but this is in some sense too blunt of a tool for our purposes: for Weyl semimetals, a more direct (co)homology interpretation in terms of Weyl points and their concomitant Dirac strings is more computationally tractable, while also offering more direct physical intuition.

In the physical setting, equivariant (co)homology is used to encode internal symmetries of the Brillouin torus $\TT^3$. Most symmetries of interest to us have a free action, i.e.\ they have no specific fixed momenta. This allows us to exploit the reduction to ordinary (co)homology in Eq.~\eqref{eq:equivariant-homology} by moving to the reduced Brillouin zone; for example, the glide symmetry in Eq.~\eqref{eq:3DGlide} has a $\ZZ_2$ group action on $\TT^3$, and the quotient space is the aforementioned $\KS$, so that e.g.\ the second cohomology groups are related as follows:
\begin{equation}
    H_{\ZZ_2}^2\left(\TT^3\right) \cong H^2\left(\KS\right).
\end{equation}
In other words, the move to a fully reduced Brillouin zone under a free action is completely justified from a (co)homology perspective (in fact, this is precisely what justifies the use of the Brillouin torus instead of translation-equivariant (co)homology in the first place). Still, the equivalent notion of equivariant (co)homology on the full Brillouin torus is often more physically insightful, since a sense of orientation is maintained on $\TT^3$. In this light, we will often take the liberty of conceptualising groups like $H^2\left(\KS\right)$ as classifying symmetric structures on the full torus.

\begin{figure}[htb!]
	\centering
	\includegraphics[width=.9\linewidth]{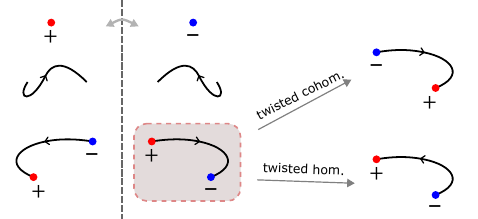}
	\caption{Schematic illustrating the necessity of twisted (co)homology groups in describing systems with orientation-reversing symmetries. 
    The top left shows the abstract action of such a symmetry (represented here by a reflection) on elements of the ordinary second cohomology group $H^2(M\setminus W)$ (represented as a Chern number around a point in $W$) and first homology group $H_1(M,W)$ (represented as an oriented string): the elements of $H^2$ experience a sign change, while the internal orientation of the elements of $H_1$ is preserved, despite the string itself being mirrored.
    As shown on the bottom left, this combined behaviour makes these ordinary groups inadequate for describing Weyl point charges connected by Dirac strings, leading to unphysical configurations in which a Dirac string would be oriented from a positive to a negative charge; this is a physical manifestation of broken Poincar\'e duality.
    The right hand side illustrates that introducing a twist in either cohomology or homology acts as a sign change under the symmetry, recovering a physically consistent structure.
    Which of these two options describes the correct physics depends on the nature of the given symmetry: symmetric pairs of Weyl nodes with opposite (same) chirality indicate that twisted (co)homology must be used.
	}\label{fig:local_coefficients}
\end{figure}

The other necessary modification from Sec.~\ref{sec:homology-modifications} is twisted (co)homology, which plays an essential role in our description.
As mentioned in Sec~\ref{sec:WSMtophom}, the usual classification of Weyl semimetals relies heavily on the Poincar\'e duality of Eq.~\eqref{eq:Poincareduality} to ensure equivalence between the cohomology picture (which classifies invariants directly) and the homology picture (which is the natural setting for the bulk--boundary correspondence).
This duality has a natural physical interpretation: it is the statement that the orientation of (homological) Dirac strings agrees consistently with the (cohomological) chiralities of the Weyl nodes at their extremities.
Poincar\'e duality is broken in the presence of orientation-reversing symmetries, so that naive use of ordinary homology and cohomology does not retain this consistency between Dirac strings and Weyl point charges. To obtain a complete physical picture, either the homology or the cohomology groups must be twisted by introducing the local coefficients $\widetilde{\ZZ}$ reviewed in Sec.~\ref{sec:homology-modifications}. In this context, these coefficients can be thought of as inducing a sign change under the action of the orientation-reversing symmetry.
The use of either twisted homology or twisted cohomology restores the sense of duality between Dirac strings and Weyl nodes, via the twisted Poincar\'e duality of Eq.~\eqref{eq:twisted-poincare}; this is illustrated in Fig.~\ref{fig:local_coefficients}.

Twisting either homology or cohomology both restores physical consistency, but the two choices may lead to fundamentally different groups. As a result, only one of the two twists leads to the correct classification for any given orientation-reversing symmetry, the correct choice depending on the exact nature of the symmetry.
Rather than motivate this choice abstractly, we propose a straightforward but exact physical indicator: the correct twist may be inferred from the relative charges of pairs of Weyl nodes that are related by the orientation-reversing symmetry.
Since these charges are both cohomological and chiral in nature, the symmetry is expected to reverse their sign under ordinary cohomology; that is, if a symmetric pair of Weyl nodes has opposite charges, this indicates that ordinary cohomology (and, by extension, twisted homology) must be used to describe the system.
Conversely, if both Weyl nodes in the pair have the same chirality, then the correct description is given using twisted cohomology and ordinary homology. This can also be seen directly in Fig.~\ref{fig:local_coefficients}.

All orientation-reversing symmetries considered in the rest of this work [notably including the glide symmetry in Eq.~\eqref{eq:3DGlide}] will be unitary, which directly constrains the symmetric partners of Weyl nodes to be of the opposite chirality.
This can be seen from the integral in Eq.~\eqref{eq:chernnumber}: the Berry curvature in the integrand is itself invariant under unitary transformations of the Hamiltonian~\cite{Kobe1990}, whereas a reversed orientation of the integration domain $S^2$ leads to a single sign change.
It follows that all unitary orientation-reversing symmetries lead to classification by ordinary cohomology and twisted homology. We will refer to the Dirac strings and loops in these twisted homology groups as \emph{twisted Dirac strings} and \emph{twisted Dirac loops}, to emphasise that they do not admit a consistent orientation within a non-orientable Brillouin zone. An example of this can be seen in Figs.~\ref{subfig:BZ_mid} and \ref{subfig:BZ_right}, where a Dirac string appears to reverse orientation from the perspective of the fundamental domain; this is a direct manifestation of the sign change induced by the local coefficients.

In contrast to this, anti-unitary orientation-reversing symmetries may in some cases give rise to same-charge symmetric pairs of Weyl nodes. Most notably, this is the case under time reversal symmetry, which induces an orientation-reversing inversion on the Brillouin zone; as such, time-reversal symmetric semimetals are properly classified using twisted cohomology and untwisted homology -- a result which has been previously established using a more abstract line of reasoning~\cite{Thiang2017}.

As a final note, we must stress that the formalism laid out above only provides a complete classification in the case when the orientation-reversing symmetry has a free action; if instead the action has fixed momenta, then additional arguments must be given to determine the role that those momenta play in the (co)homology classification.
We explore a heuristic approach to such arguments in the important case of inversion-symmetric Weyl semimetals in Sec.~\ref{sec:invsym}.

\section{Topology of \texorpdfstring{$\KS$}{K²×S¹}} \label{sec:k2s1}
We now specify the formalism laid out in the previous section to the glide symmetry in Eq.~\eqref{eq:3DGlide}, to obtain a novel (co)homological description of the Weyl semimetals with $\KS$ Brillouin zone appearing in Ref.~\cite{fonseca2024}.
As explained in Sec.~\ref{sec:motivation}, the free action of this glide symmetry justifies the use of (co)homology on the quotient space $\KS$ in place of its equivariant version on the full torus $\TT^3$. Furthermore, the unitary nature of the symmetry indicates the use of twisted homology and ordinary cohomology.
As a result, we find that the topology of this system is classified using the following Mayer--Vietoris sequence, where we notate $M\coloneqq\KS$:
\begin{equation}\label{eq:MV-nonorientable}
	0\ \to\ \underbrace{H^2(M)}_{\mathclap{\text{Insulator}}}\ \to\ \underbrace{H^2\left(M\setminus W\right)}_{\mathclap{\text{Semimetal}}}\ \to\ \underbrace{\bigoplus_{w\in W} H^2\left(S_w^2\right)}_{\text{Local charges}}\ \overset{\Sigma}{\to}\ \underbrace{H^3\left(M\right)}_{\mathclap{\text{Total charge}}}\ \to\ 0.
\end{equation}
Equivalently, the Poincar\'e dual sequence in twisted homology groups reads
\begin{equation}\label{eq:homology-sequence-nonorientable}
	0\ \to\ \underbrace{H_1\left(M;\ZZtwist\right)}_{\mathclap{\text{Twisted Dirac loops\hspace{1em}}}}\ \to\ \underbrace{H_1\left(M, W;\ZZtwist\right)}_{\mathclap{\text{\hspace{1em}Twisted Dirac strings}}}\ \overset{\partial}{\to}\ H_0\left(W;\ZZtwist\right)\ \overset{\Sigma}{\to}\ H_0\left(M;\ZZtwist\right)\ \to\ 0.
\end{equation}
Since Eq.~\eqref{eq:MV-nonorientable} is given in terms of ordinary cohomology groups, it can be readily calculated using standard tools in algebraic topology, for instance employing cellular homology~\cite{Hatcher}.
The sequences explicitly read
\begin{equation}\label{eq:explicit-sequence-nonorientable}
	0\to \ZZ\oplus\ZZ_2 \overset{\alpha}{\to} \ZZ\oplus\ZZ_2\oplus \ZZ^k \overset{\beta}{\to} \ZZ^k \overset{\Sigma}{\to} \ZZ_2 \to 0,
\end{equation}
with $k = |W|$ being the total number of Weyl points and the map $\alpha$ the inclusion map (in the sense that each insulator corresponds to a semimetal with all Weyl point charges set to zero).
In what follows, we discuss the topological and physical consequences of this sequence.

\subsection{Mod 2 charge cancellation} \label{sec:mod2cc}
Before explicitly discussing the topological invariants and their nature, we will address the (co)homological version of the main result of Ref.~\cite{fonseca2024}, namely the mod 2 charge cancellation condition.
This condition relates directly to the fact that the rightmost group in the sequence Eq.~\eqref{eq:explicit-sequence-nonorientable} is $\ZZ_2$. This is a completely general feature for non-orientable manifolds; the group appearing here is the top cohomology group, and $H^n(M)=\ZZ_2$ for any $n$-dimensional non-orientable manifold $M$.

Recall that in the orientable case, charge cancellation is obtained from the sequence in Eq.~\eqref{eq:semimetal-MV-explicit} by noting that exactness forces the Weyl point charges in $\ZZ^k$ to sum to $0\in\ZZ$ whenever they descend from a semimetal phase.
The reasoning is similar in the non-orientable case, but one needs to be a bit more careful in the sense that the group $\ZZ^k$ appearing in Eq.~\eqref{eq:explicit-sequence-nonorientable} no longer admits a canonical basis.
This can be understood from both the homology and the cohomology perspective.

On the homology side, the map $\beta$ in Eq.~\eqref{eq:explicit-sequence-nonorientable} is the boundary map $\partial$ from Eq.~\eqref{eq:homology-sequence-nonorientable}. 
Ordinarily, this map assigns opposite signs to both ends of a Dirac string, giving immediate charge cancellation. 
However, the use of twisted homology complicates this story, since it induces sign changes along orientation reversing paths. 
As such, the relative sign between the endpoints is only well defined when their relative orientation is, i.e.\ when the Weyl points coincide; in these cases, the Dirac string forms a closed loop in $\KS$, and the relative sign determines whether the Weyl points are allowed to annihilate. 
In particular, the Weyl points have the same sign whenever this loop corresponds to the $\ZZ_2$ generator of the fundamental group $\pi_1(\KS)$, which in the language of Ref.~\cite{fonseca2024} is equivalent to crossing the orientation reversing boundary of the fundamental domain an odd number of times.

In cohomology terms, $\beta$ should be thought of as the restriction of the semimetallic first Chern class in $H^2(M-W)$ to the spheres surrounding the Weyl points.
Importantly, the Chern classes in $H^2\left(\bigcup_{i=1}^k S^2_{w_i}\right) \cong \ZZ^k$ can no longer be directly identified with Weyl point charges, since this requires a canonical orientation on each of the spheres $S^2_{w_i}$.
Indeed, fixing a basis for this group is equivalent to fixing a specific orientation at each Weyl point, a choice that is directly related to the freedom to choose different fundamental domains in the manner illustrated in Fig.~\ref{fig:BZ_param}.
Importantly, however, the map $\Sigma$ acts as a well-defined sum of the entries in this $\ZZ^k$, independently of this choice of basis: precisely because it maps onto $\ZZ_2$, it is invariant under changes of orientation ($\chi\equiv-\chi \mod{2}$).
Therefore, we can write
\begin{equation} \label{eq:mod2cc}
    \Sigma: \ZZ^k \to \ZZ_2, \quad (\chi_1,...,\chi_k)\mapsto \sum_{i=1}^k \chi_i \mod{2}
\end{equation}
without fixing the sign of each $\chi_i$ in advance.

With this knowledge in hand, the argument proceeds similarly to the orientable case: since the sequence is exact, charges initially belonging to a semimetal structure [which are therefore mapped into $\ZZ^k$ by the map $\beta$ in Eq.~\eqref{eq:explicit-sequence-nonorientable}] necessarily add to $0\in \ZZ_2$ in any basis.
This is the precise coordinate-free interpretation of the mod 2 charge cancellation theorem presented in Eq.~\eqref{eq:mod2cc}, stemming from Ref.~\cite{fonseca2024}.
Importantly, note that our construction imposes meaningful limits on the interpretation of this charge cancellation: on the one hand, the lack of a canonical basis represents an unphysical degree of freedom, implying that there is no physical interpretation in terms of a particular non-zero total charge; on the other hand, the appearance of $H^3(\KS)\cong\ZZ_2$ ensures that such an interpretation is also not mathematically meaningful, since the total charge is fundamentally a $\ZZ_2$ invariant.

Beyond this charge cancellation theorem, the exactness of the sequence Eq.~\eqref{eq:explicit-sequence-nonorientable} also allows us to conclude the converse: if any charge configuration in $\ZZ^k$ adds to $0\in \ZZ_2$, it necessarily admits a semimetal configuration. Concretely, this means that any combination of charges inside a given fundamental domain that has an even total can be realised in a Weyl semimetal phase.

\subsection{Topological invariants} \label{sec:TIs}

As mentioned in Secs.~\ref{sec:WSMtopcohom} and \ref{sec:WSMtophom}, Weyl semimetals can be considered transitional phases between different insulating phases, and their topology can be understood in terms of insulating invariants. As such, understanding the set of semimetal invariants on $\KS$ begins with a good grasp of the insulating invariants.
These invariants are classified by the first group appearing in Eqs.~\eqref{eq:MV-nonorientable}, \eqref{eq:homology-sequence-nonorientable} and \eqref{eq:explicit-sequence-nonorientable}, namely the group
\begin{equation}
    H^2\left(\KS\right) \cong H_1\left(\KS;\ZZtwist\right) \cong \ZZ\oplus\ZZ_2
\end{equation}
In contrast to the conventional three-dimensional insulating topology, which is given by three Chern numbers $C_i\in\ZZ$ stemming from $H^2\left(\TT^3\right)\cong \ZZ^3$, the insulating phases on $\KS$ are classified by just two invariants: one $\ZZ$ invariant $\nu_x$ and one $\ZZ_2$ invariant $\nu_z$.
Alternatively, the topology may be compared to that of insulating systems subject to an ordinary reflection symmetry: in this case, the topological invariant is reduced to a single $\ZZ$-valued Chern number in the direction of the mirror plane~\cite{Hsieh2012,Chiu2013}. 
In this light, going from an ordinary reflection to a glide symmetry induces an additional $\ZZ_2$-valued invariant in the direction unaffected by the symmetry operation.

\begin{figure}[htb!]
	\centering
	\subcaptionbox{$[\tilde{\ell}_x]$\label{subfig:Z_invariant}} {\includegraphics[width=.3\textwidth]{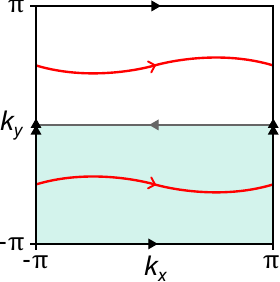}}
	\hfil
	\subcaptionbox{$[\tilde{\ell}_y] = 0$\label{subfig:0_invariant}} {\includegraphics[width=.3\textwidth]{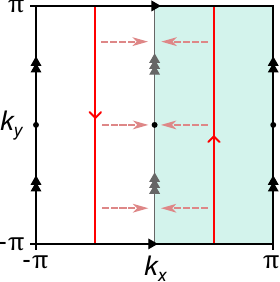}}
	\hfil
	\subcaptionbox{$[\tilde{\ell}_z] = [-\tilde{\ell}_z]$\label{subfig:Z2_invariant}} {\includegraphics[width=.3\textwidth]{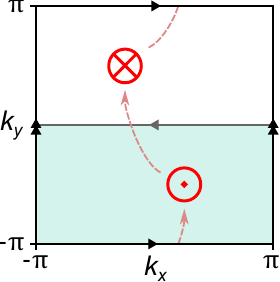}}
	\caption{Top view of the $\TT^3$ Brillouin torus. Each of the three basic Dirac loops $\ell_i$ are shown contained within an appropriately chosen $\KS$ fundamental domain (shaded teal), and doubled by glide symmetry to the other half of the torus. The combined structure (shown in red) can be regarded as a twisted Dirac loop $\tilde{\ell}_i$ [representing an invariant in $H_1(\KS;\ZZtwist)$] in each instance. 
    (a) For $\ell_x$, the parity change induced by $k_x \mapsto -k_x$ is cancelled out by the orientation reversal induced by the twist in the homology. The resulting twisted Dirac string $\tilde{\ell}_x$ thus has a consistent orientation, and the associated invariant $\nu_x\in\ZZ$ induces an even Chern number $C_x = 2\nu_x\in 2\ZZ$ on the full torus. 
    (b) For the case of $\ell_y$, the corresponding twisted Dirac loop $\tilde{\ell}_y$ is trivial: the mirroring along $k_x=0$ and subsequent orientation reversal result in a set of two oppositely orientated loops. These loops can be brought together along the dashed arrows in a symmetry-preserving fashion, eventually cancelling when they meet on the glide plane. As a result, there is no invariant $\nu_y$.
    (c) The Dirac loop $\ell_z$ running in the positive $k_z$ direction (``out of the page'') in the fundamental domain is doubled to a second loop running in the negative $k_z$ direction (``into the page''). These loops cannot be merged in a fashion consistent with the symmetry, but they can instead be permuted along the dashed arrows. As a consequence, the resulting twisted Dirac string $\tilde{\ell}_z$ is equivalent to its own inverse $-\tilde{\ell}_z$; that is, $\tilde{\ell}_z$ generates a $\ZZ_2$ invariant $\nu_z$.}
	\label{fig:K2S1_invariants} 
\end{figure}

These invariants can be most intuitively understood by studying how the glide symmetry acts on the Dirac loops in $H_1\left(\TT^3\right)$ to give twisted Dirac loops in $H_1\left(\KS;\ZZtwist\right)$.
In the absence of any symmetries, the insulating homology group $H_1\left(\TT^3\right)\cong\ZZ^3$ is generated by the homology classes of the three basic Dirac loops spanning the different coordinate directions of the Brillouin torus; we denote these loops by $\ell_i$ with $i\in\{x,y,z\}$. As explained in Sec.~\ref{sec:WSMtophom}, these loops are precisely dual to the three insulating Chern numbers $\vb{C}=\left(C_x,C_y,C_z\right)$ defined in Eq.~\eqref{eq:insulatorchern}.
The glide symmetry induces a symmetric partner $\ell'_i$ to each of these loops, and their respective sums $\tilde{\ell}_i=\ell_i+\ell'_i$ represent elements in the twisted homology group $H_1\left(\KS;\ZZtwist\right)$, due to the equivalence between homology on $\KS$ and equivariant homology on $\TT^3$ mentioned in Sec.~\ref{sec:motivation}; this construction is illustrated in Fig.~\ref{fig:K2S1_invariants}.
Together, these three twisted Dirac loops generate the two invariants mentioned above: $\tilde{\ell}_x$ generates the $\ZZ$ invariant $\nu_x$, $\tilde{\ell}_y$ is topologically trivial, and $\tilde{\ell}_z$ generates the $\ZZ_2$ invariant $\nu_z$.
Of these, the $\ZZ_2$ invariant $\nu_z$ is closely related to the $\ZZ_2$ invariant found on the two-dimensional Klein bottle $K^2$ in Ref.~\cite{Chen2022}, and its appearance and properties in the semimetallic $\KS$ context has been discussed in detail in Ref.~\cite{fonseca2024}.

The $\ZZ$ invariant $\nu_x$ has not been previously reported; in the insulating context it can be naively calculated as half of the Chern number $C_x$ appearing in Eq.~\eqref{eq:insulatorchern}:
\begin{equation} \label{eq:topinvx}
\nu_x = \frac{1}{2}C_x = \frac{1}{4\pi}\int_{\TT^2_{k_x=\text{const.}}} \mathcal{F}.
\end{equation}
This integral, despite not necessarily respecting the glide symmetry, yields a consistent value because the symmetry precisely doubles $\ell_x$-like Dirac loops.

In the presence of Weyl nodes, the group of invariants is instead given by the semimetal group
\begin{equation} \label{eq:semimetal-group-nonor}
    H^2\left(\KS\setminus W\right) \cong H_1\left(\KS,W;\ZZtwist\right) \cong \ZZ\oplus \ZZ_2 \oplus \ZZ^k,
\end{equation}
where $k=|W|$ is the number of Weyl points.
The structure of this group is reminiscent of the $\ZZ^3\oplus\ZZ^{k-1}$ group appearing in the orientable case in Eq.~\eqref{eq:semimetal-MV-explicit}, in that it combines the insulating topology with a set of Weyl point charge configurations.
The one crucial difference is that these charge configurations are represented by a $\ZZ^k$ subgroup rather than $\ZZ^{k-1}$, reflecting the weaker mod 2 charge cancellation condition.
Formally, this is understood by using the exactness of the sequence in Eq.~\eqref{eq:explicit-sequence-nonorientable}, which gives
\begin{equation}
    \text{im}\left(\beta\right) = \text{ker}\left(\Sigma\right) = \bigoplus_{w\in W} H^2\left(S_w^2\right)/H^3\left(\KS\right) \cong \ZZ^k/\ZZ_2 \cong \ZZ^k.
\end{equation}
This is exactly the $\ZZ^k$-factor appearing in Eq.~\eqref{eq:explicit-sequence-nonorientable}, but, crucially, this should not be directly interpreted directly as the charged of the $k$ Weyl points, but rather only those configurations summing to $0$ in $\ZZ_2$.
A similar line of reasoning holds in the orientable case, where the conventional charge cancellation theorem results in the corresponding $\ZZ^{k-1}$-factor.~\cite{Mathai2017a,Mathai2017b}
Physically, this implies that a single Weyl point on $\KS$ may give rise to valid semimetal topology, as opposed to the minimum of two Weyl points that are necessary in the orientable case.
Such a lone Weyl point necessarily has an even charge, enforced by the mod 2 charge cancellation; this was also pointed out in the supplement to Ref.~\cite{fonseca2024}.

\begin{figure}[htb!]
	\centering
	\def\svgwidth{.85\linewidth}
\begingroup%
  \makeatletter%
  \providecommand\color[2][]{%
    \errmessage{(Inkscape) Color is used for the text in Inkscape, but the package 'color.sty' is not loaded}%
    \renewcommand\color[2][]{}%
  }%
  \providecommand\transparent[1]{%
    \errmessage{(Inkscape) Transparency is used (non-zero) for the text in Inkscape, but the package 'transparent.sty' is not loaded}%
    \renewcommand\transparent[1]{}%
  }%
  \providecommand\rotatebox[2]{#2}%
  \newcommand*\fsize{\dimexpr\f@size pt\relax}%
  \newcommand*\lineheight[1]{\fontsize{\fsize}{#1\fsize}\selectfont}%
  \ifx\svgwidth\undefined%
    \setlength{\unitlength}{336.65684257bp}%
    \ifx\svgscale\undefined%
      \relax%
    \else%
      \setlength{\unitlength}{\unitlength * \real{\svgscale}}%
    \fi%
  \else%
    \setlength{\unitlength}{\svgwidth}%
  \fi%
  \global\let\svgwidth\undefined%
  \global\let\svgscale\undefined%
  \makeatother%
  \begin{picture}(1,0.39806837)%
    \lineheight{1}%
    \setlength\tabcolsep{0pt}%
    \put(0,0){\includegraphics[width=\unitlength,page=1]{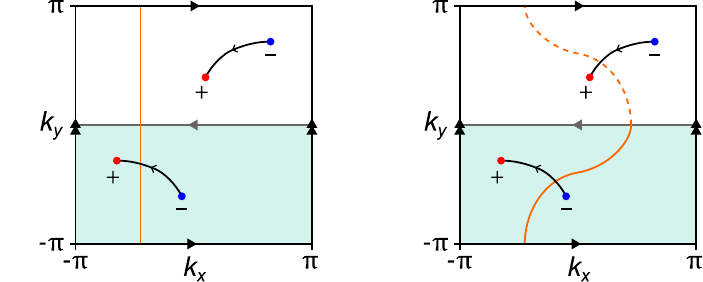}}%
    \put(0.54532074,0.35596869){\color[rgb]{0,0,0}\makebox(0,0)[lt]{\lineheight{1.25}\smash{\begin{tabular}[t]{l}(b)\end{tabular}}}}%
    \put(-0.00197832,0.35596869){\color[rgb]{0,0,0}\makebox(0,0)[lt]{\lineheight{1.25}\smash{\begin{tabular}[t]{l}(a)\end{tabular}}}}%
  \end{picture}%
\endgroup%

	\caption{Incorrect (a) and correct (b) calculations of the $\nu_x$ invariant in a given setting. 
    (a) In a given semimetal, the slice of constant $k_x$ (orange) appearing in Eq.~\eqref{eq:topinvx} may only intersect a single Dirac string, yielding the non-integer value of 1/2 in the pictured setup. (b) If integration is instead carried out over a topologically equivalent glide symmetric surface, the topological invariant $\nu_x$ is well-defined outside of Weyl points. The dashed half of the surface is redundant, since it yields the same integral as the solid half. In other words, to calculate the invariant $\nu_x$ in a semimetallic setting, the integration surface must be a closed/compact submanifold of $\KS$, which explains why the configuration in (b) yields a consistent invariant while the constant $k_x$-slice in (a) does not. The pictured setup yields $\nu_x=1$. Note that the configuration of the Weyl nodes, Dirac strings and their respective symmetric partners above indicates the use of twisted homology; cf. Fig.~\ref{fig:local_coefficients}.}
	\label{fig:nu-x}
\end{figure}

Similar to conventional Weyl semimetals, the Weyl points serve as singularities of the Berry field, within any given choice of the fundamental domain. As a result, the invariant on a two-dimensional slice of the Brillouin zone is modified by an amount dictated by the Weyl point charge.
This modification has to be treated with care in the non-orientable case, in a way which respects the glide symmetry on the Brillouin zone.
This is true in particular in the case of $\nu_x$; as mentioned above, the integral in  Eq.~\eqref{eq:topinvx} does not necessarily respect the glide symmetry. This does not lead to any issues in the insulating case, but as illustrated in Fig.~\ref{fig:nu-x}(a), it may lead to unphysical results in the semimetallic case.
Along the same line as for time-reversal symmetric systems, where integrals have to be taken over $\mathcal{T}$-invariant surfaces~\cite{Thiang2017}, the integral instead needs to be taken over a surface $S$ that preserves the glide symmetry and is continuously deformable into a surface of constant $k_x$; see Fig.~\ref{fig:nu-x}(b).
In principle, only the half of the surface $S$ that intersects the fundamental domain at $k_y\leq 0$ needs to be integrated over; this half can be considered a compact subspace of $\KS$.
The explicit calculation can thus be written as
\begin{equation}
    \nu_x = \frac{1}{2\pi} \int_{S_{1/2}} \mathcal{F},
\end{equation}
where $S_{1/2}$ denotes the appropriate half of the surface $S$.
This integration domain may be continuously deformed while respecting the symmetry.
As alluded to above, passing it over a Weyl point with charge $\chi$ alters $\nu_x$ to a new value of $\nu_x' = \nu_x\pm\chi$, with the choice of sign depending on the direction in which the Weyl point passes through the surface.

The $\ZZ_2$ invariant $\nu_z$ is somewhat more lenient with respect to the choice of surface: since the glide symmetric partner of a Weyl node always has the same $k_z$-value, this invariant may consistently be calculated on slices of constant $k_z$, as is done in Ref.~\cite{fonseca2024}.
This invariant obeys a similar rule to $\nu_x$ with respect to different integration domains: letting a constant-$k_z$ slice with associated $\ZZ_2$-invariant $\nu_z$ pass over a Weyl point with charge $\chi$ along the $k_z$-direction, the new $\ZZ_2$-invariant becomes $\nu_z' = \nu_z+\chi \mod{2}$.
It should be emphasised that no $\ZZ_2$ charge needs to be assigned to each individual Weyl point to support this behaviour, contrary to the description in Ref.~\cite{fonseca2024}.
This is reflected in the structure of the semimetal group in Eq.~\eqref{eq:semimetal-group-nonor}: the only $\ZZ_2$ factor in this group stems from the underlying insulating topology, and the addition of any Weyl points only contributes additional $\ZZ$ factors.
That is, no $\ZZ_2$ topology is intrinsically associated with the Weyl point charge.
A comparable construction appears in the case of time-reversal symmetric Weyl semimetals: there, Weyl points also mediate a $\ZZ_2$ invariant without having intrinsic $\ZZ_2$ topology~\cite{Thiang2017}.

\subsection{\texorpdfstring{$K^2$}{K²} surface states} \label{sec:k2surface}

Having focused exclusively on the description of the bulk topology above, we now turn our attention to the classification of the corresponding boundary states on the non-orientable $(k_x,k_y)$-surface.
These states are obtained by performing a projection $\pi_z$ along the periodic $k_z$ direction, effectively integrating out the $S^1$ factor in $\KS$; the resulting surface Brillouin zone then has the topology of $K^2$.
Just as in the bulk, the lack of a consistent choice of orientation leads the chirality of the surface states (i.e.\ the Fermi arcs) to be ill-defined; as such, a coordinate-free description offers similar insights in this setting.

The description of surface topology is obtained by letting the projection $\pi_z$ act on homology classes, analogous to the construction in Sec.~\ref{sec:bulkboundary}.
Concretely, any given twisted relative homology class $[c]\in H_1\left(\KS,W,\ZZtwist\right)$ is represented by a configuration of twisted Dirac strings $c$.
The surface projection $\pi_z$ naturally preserves the twisted orientation of such configurations, allowing a well-defined pushforward $\left(\pi_z\right)_*$ on twisted homology classes:
\begin{equation}
    \left(\pi_z\right)_* : H_1\left(\KS,W;\ZZtwist\right) \to H_1\left(K^2,W';\ZZtwist\right), \quad [c]\mapsto \left[\pi_z(c)\right],
\end{equation}
where $W'$ denotes the projected surface Weyl nodes as before.
Similarly to the orientable case, these surface homology classes are exactly represented by the physical Fermi arcs.
Given the correct orientation, it is therefore appropriate to think of these structures in terms of \emph{twisted Fermi arcs}.
Furthermore, the pushforward $\left(\pi_z\right)_*$ is a surjective map, ensuring that all physically realisable Fermi arc configurations on $K^2$ are classified by the group $H_1\left(K^2,W';\ZZtwist\right)$.

Just as in the orientable case, the pushforward construction may be extended to the entire homology sequence in Eq.~\eqref{eq:homology-sequence-nonorientable}, yielding the following exact sequence on the surface:
\begin{equation}
	0\ \to\ \underbrace{H_1\left(K^2;\ZZtwist\right)}_{\mathclap{\text{Twisted Fermi loops}}}\ \to\ \underbrace{H_1\left(K^2, W';\ZZtwist\right)}_{\mathclap{\text{Twisted Fermi arcs}}}\ \overset{\partial}{\to}\ H_0\left(W';\ZZtwist\right)\ \to\ H_0\left(K^2;\ZZtwist\right)\ \to\ 0.
\end{equation}
For calculational purposes, it is beneficial to employ twisted Poincar\'e duality on the surface in order to recover a Mayer--Vietoris sequence in terms of more tractable untwisted cohomology groups:
\begin{equation}\label{eq:K2-surface-cohomology}
	\cdots\ \overset{0}{\to}\ H^1\left(K^2\right)\ \to\ H^1\left(K^2\setminus W'\right)\ \to\ \bigoplus_{w'\in W'} H^1\left(S_{w'}^1\right)\ \to\ H^2\left(K^2\right)\ \to\ 0.
\end{equation}
Again, the use of first rather than second cohomology signifies that the surface invariants are calculated in terms of winding numbers of a Berry connection 1-form, rather than Chern numbers.
Just as in Sec.~\ref{sec:bulkboundary}, we emphasize that this Berry connection 1-form is calculated using a two-dimensional projected Bloch Hamiltonian, rather than on the boundary of a Hamiltonian with open boundary conditions~\cite{Mathai2017b}. 
This projected Hamiltonian is not gapless outside of the projected Weyl points, allowing for consistent calculation of Berry phases.

Just as in the bulk, the groups in Eq.~\eqref{eq:K2-surface-cohomology} can be straightforwardly calculated using e.g.\ cellular homology, resulting in the following explicit sequence:
\begin{equation}\label{eq:K2-MV-explicit}
	\cdots \overset{0}{\to} \ZZ \to \ZZ\oplus\ZZ^k \to \ZZ^k \overset{\Sigma}{\to} \ZZ_2 \to 0.
\end{equation}
This sequence is more or less identical in form and interpretation to the bulk sequence in Eq.~\eqref{eq:explicit-sequence-nonorientable}; the only essential difference is that the $\ZZ_2$-factor corresponding to the invariant $\nu_z$ is missing as a consequence of projecting along the $k_z$-direction.
As a result, the $K^2$ surface is subject to analogous mod 2 charge cancellation, admitting the same interpretation as in Sec.~\ref{sec:mod2cc}.
In particular, the lack of a canonical basis for $H^1\left(K^2\setminus W'\right)\cong\ZZ^k$ implies that the chirality of individual surface Weyl nodes is fundamentally ill-defined, as is the relative chirality between them.
Physically, this means that different choices of the fundamental domain can be used to obtain seemingly different Fermi arc structures. 

\begin{figure}[htb!]
	\centering
	\includegraphics[width=.45\linewidth]{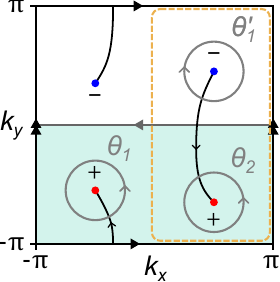}
	\caption{Schematic contextualisation of the experiment carried out in Ref.~\cite{fonseca2024}, where the surface dispersion is measured along the loops $\theta_1$ and $\theta_2$ encircling the projections of two positive Weyl nodes onto a $K^2$ fundamental domain (shaded teal). This measurement yields two surface states of the same apparent chirality. If one instead chooses the alternative fundamental domain outlined in orange, the loop $\theta_1$ is mirrored to a different loop $\theta_1'$ by the glide symmetry, which encircles the projection of a negative Weyl node in the clockwise direction. This change in orientation counteracts the change of chirality of the corresponding Weyl node, so that the dispersion along $\theta_1'$ is identical to that along $\theta_1$.
    Returning $\theta_1'$ to the conventional counterclockwise orientation yields the expected dispersion of a surface state with negative chirality; as a result, the alternative fundamental domain will be found to feature two oppositely charged points connected by a Fermi arc displaying conventional dispersion characteristics.}
	\label{fig:K2-experiment-alt}
\end{figure}

This ambiguity helps explain the experimental results in Ref.~\cite{fonseca2024}, where two surface Weyl nodes are found to have the same chirality with respect to a given $K^2$ fundamental domain.
Explicitly, this is probed by measuring the dispersion along two loops encircling the surface projections of two positive Weyl nodes.
In light of the results obtained here, this observation is highly dependent on the choice of fundamental domain and can be altered by including different regions of the torus; this is illustrated in more detail in Fig.~\ref{fig:K2-experiment-alt}.
Consequently, we propose that the observations made in Ref.~\cite{fonseca2024} should not be interpreted as novel physical behaviour, but as a matter of perspective; indeed, from the point of view of the full Brillouin torus $\TT^3$ the behaviour is fully conventional.

Finally, the sequence in Eq.~\eqref{eq:K2-MV-explicit} has a natural interpretation in terms of two-dimensional non-Hermitian systems; this correspondence is explored in greater depth in Sec.~\ref{sec:NH}.

\section{Extensions} \label{sec:extensions}
The formalism for topological classification under orientation-reversing symmetries introduced in Sec.~\ref{sec:motivation} is formulated in very general terms, and has many potential applications beyond the $\KS$ system studied in Sec.~\ref{sec:k2s1}.
Three such applications are studied in this section: we classify the complete set of possible non-orientable Brillouin zones and their associated invariants (Sec.~\ref{sec:nonsymsym}), discuss extensions to the topology of exceptional points in non-Hermitian systems (Sec.~\ref{sec:NH}), and use a heuristic ansatz to arrive at a classification scheme for the experimentally important class of Weyl semimetals subject to inversion symmetry (Sec.~\ref{sec:invsym}).

\subsection{Additional non-orientable Brillouin zones} \label{sec:nonsymsym}

The three-dimensional glide symmetry studied in Sec.~\ref{sec:k2s1} is not the only space group symmetry with free action resulting in a non-orientable fundamental domain.
Out of the 230 space groups in three dimensions, a total of 12 have a non-trivial free action, allowing them to induce proper reduced Brillouin zones~\cite{Michel2001}.
Four of these symmetries are orientation-reversing, in that they include at least one glide symmetry; the other eight are all generated by orientation-preserving screw rotations (i.e.\ combinations of rotation and fractional lattice translation).
These four orientation-reversing groups are precisely the ones which give rise to a non-orientable Brillouin zone. The resulting three-dimensional spaces are examples of so-called \emph{platycosms}~\cite{Conway2003}; the topology of these platycosm Brillouin zones has been treated in some generality in Ref.~\cite{Zhang2025}, especially in the insulating setting. In this section we specialise these findings to the setting of semimetallic Mayer--Vietoris sequences and argue for the utility of our twisted Dirac string formalism.

\begin{figure}[htb!]
    \centering
	\def\svgwidth{\linewidth}
\begingroup%
  \makeatletter%
  \providecommand\color[2][]{%
    \errmessage{(Inkscape) Color is used for the text in Inkscape, but the package 'color.sty' is not loaded}%
    \renewcommand\color[2][]{}%
  }%
  \providecommand\transparent[1]{%
    \errmessage{(Inkscape) Transparency is used (non-zero) for the text in Inkscape, but the package 'transparent.sty' is not loaded}%
    \renewcommand\transparent[1]{}%
  }%
  \providecommand\rotatebox[2]{#2}%
  \newcommand*\fsize{\dimexpr\f@size pt\relax}%
  \newcommand*\lineheight[1]{\fontsize{\fsize}{#1\fsize}\selectfont}%
  \ifx\svgwidth\undefined%
    \setlength{\unitlength}{445.2406275bp}%
    \ifx\svgscale\undefined%
      \relax%
    \else%
      \setlength{\unitlength}{\unitlength * \real{\svgscale}}%
    \fi%
  \else%
    \setlength{\unitlength}{\svgwidth}%
  \fi%
  \global\let\svgwidth\undefined%
  \global\let\svgscale\undefined%
  \makeatother%
  \begin{picture}(1,0.36931327)%
    \lineheight{1}%
    \setlength\tabcolsep{0pt}%
    \put(0.70713923,0.33918828){\color[rgb]{0,0,0}\makebox(0,0)[lt]{\lineheight{1.25}\smash{\begin{tabular}[t]{l}(c) $Pna2_1$\end{tabular}}}}%
    \put(0.00681899,0.33918828){\color[rgb]{0,0,0}\makebox(0,0)[lt]{\lineheight{1.25}\smash{\begin{tabular}[t]{l}(a) $Cc$\end{tabular}}}}%
    \put(0.35697911,0.33918828){\color[rgb]{0,0,0}\makebox(0,0)[lt]{\lineheight{1.25}\smash{\begin{tabular}[t]{l}(b)  $Pca2_1$\end{tabular}}}}%
    \put(0,0){\includegraphics[width=\unitlength,page=1]{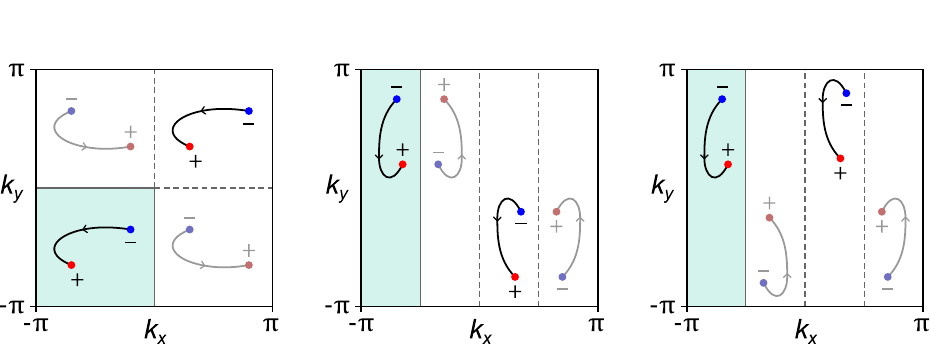}}%
  \end{picture}%
\endgroup%

    \caption{Top view of the Brillouin torus $\TT^3$, subject to the space group symmetries (a) $Cc$, (b) $Pca2_1$, and (c) $Pna2_1$. In each case, an arbitrary choice of fundamental domain is given (shaded teal). To help illustrate the action of each symmetry, a structure consisting of a Dirac string connecting two Weyl nodes is placed in this fundamental domain; three symmetric copies of this structure then appear in other sections of the Brillouin torus. Greyed-out structures indicate copies that have been translated by $\pi$ in the $k_z$-direction with respect to the original, and thus do not lie on the same $k_z$-plane.
    The fundamental domains shown are made into non-orientable spaces by identifying the elements of their boundary which are related by the symmetry: for example, the $k_y=0$ face of the fundamental domain in (a) has an internal symmetry by Eq.~\eqref{eq:Cc_constraint1}, which identifies the half at $-\pi\leq k_z\leq 0$ with that at $0\leq k_z\leq\pi$. This can also be seen directly: if a Weyl point in this fundamental domain is taken across the $k_y=0$ boundary, its (greyed-out) symmetric partner emerges at the same point translated by $\pi$ in the $k_z$-direction.
    These identifications give rise to cell structures, which can be used to calculate groups of invariants via cellular homology. Importantly, these groups are independent of the choice of fundamental domain.}
    \label{fig:alternative_BZs}
\end{figure}
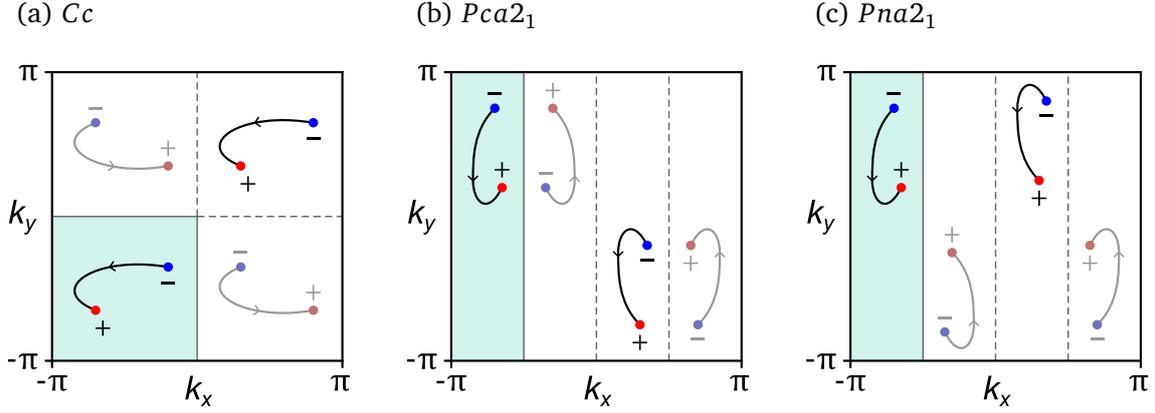

The simplest of these four groups is $Pc$, which is generated by a single glide symmetry. The resulting platycosm is the \emph{first amphicosm}, which is precisely the space $K²\times S^1$ treated in detail in Sec.~\ref{sec:k2s1} above.
The remaining three groups are $Cc$, $Pca2_1$, and $Pna2_1$, and each of them combines a glide symmetry with another $\ZZ_2$ operation.
As a consequence, these three groups are homomorphic to $\ZZ_2\oplus \ZZ_2$, as opposed to the $\ZZ_2$ corresponding to $Pc$. 
It follows that their respective fundamental domains cover a quarter of the Brillouin torus rather than half of it; see Fig.~\ref{fig:alternative_BZs}.
Just as in the case of $\KS$, the platycosm topology is obtained when their boundaries are identified according to the action of the symmetry.
Notably, these boundary identifications naturally induce cell structures on the fundamental domain, which can be used to calculate the relevant exact sequences of invariants using cellular homology.

The space group symmetries under consideration are unitary in nature; as discussed in Sec.~\ref{sec:motivation}, this ensures that the correct topological description is given in terms of ordinary cohomology and twisted homology.
For each symmetry, we list the precise action on the momentum-space Hamiltonian, together with the explicitly calculated bulk semimetal Mayer--Vietoris sequence as it appears in Eq.~\eqref{eq:MV-nonorientable}. $U_X$ will denote unitary representations of the symmetries in the space group $X$.

First, the $Cc$ group combines a glide symmetry and a diagonal half-lattice translation, resulting in a Brillouin zone topology called the \emph{second amphicosm}. The symmetry constraints on the Hamiltonian read
\begin{align}
	\Ham\left(k_x, k_y, k_z\right) &= U_{Cc_1}\Ham\left(k_x, -k_y, k_z+\pi\right)U^{-1}_{Cc_1}, \label{eq:Cc_constraint1}\\
	\Ham\left(k_x, k_y, k_z\right) &= U_{Cc_2}\Ham\left(k_x+\pi, k_y+\pi, k_z\right)U^{-1}_{Cc_2}.
\end{align}
The bulk Mayer--Vietoris sequence in ordinary cohomology (equivalently, the exact sequence in twisted homology) is as follows:
\begin{equation} \label{eq:NS1}
	0 \to \ZZ \to \ZZ\oplus\ZZ^k \to \ZZ^k \to \ZZ_2 \to 0.
\end{equation}
Remarkably, the insulating group $\ZZ$ does not include any finite-order invariants; in particular, compared to the  $\ZZ\oplus\ZZ_2$ group on $\KS$, the half-lattice translation prevents a non-trivial $\ZZ_2$ invariant from appearing.

Second, the group $Pca2_1$ results in the \emph{first amphidicosm}; this is generated by a combined glide reflection and screw rotation, which manifest in momentum space as
\begin{align}
	\Ham\left(k_x, k_y, k_z\right) &= U_{\left(Pca2_1\right)_1}\Ham\left(k_x+\pi, -k_y, k_z\right)U^{-1}_{\left(Pca2_1\right)_1}, \\
	\Ham\left(k_x, k_y, k_z\right) &= U_{\left(Pca2_1\right)_2}\Ham\left(-k_x, -k_y, k_z+\pi\right)U^{-1}_{\left(Pca2_1\right)_2},
\end{align}
and give rise to the following bulk sequence:
\begin{equation} \label{eq:NS2}
	0 \to \ZZ_2^2 \to \ZZ_2^2\oplus\ZZ^k \to \ZZ^k \to \ZZ_2 \to 0.
\end{equation}

Finally, the $Pna2_1$ group is formed by operations similar to those defining $Pca2_1$, but with an offset in the glide reflection plane from $k_y=0$ to $k_y=\pm\pi/2$.
The resulting space is called the \emph{second amphidicosm}, and the corresponding symmetry relations are
\begin{align}
	\Ham\left(k_x, k_y, k_z\right) &= U_{\left(Pna2_1\right)_1}\Ham\left(k_x+\pi, \pi-k_y, k_z\right)U^{-1}_{\left(Pna2_1\right)_1}, \\
	\Ham\left(k_x, k_y, k_z\right) &= U_{\left(Pna2_1\right)_2}\Ham\left(-k_x, -k_y, k_z+\pi\right)U^{-1}_{\left(Pna2_1\right)_2},
\end{align}
with a bulk Mayer--Vietoris sequence of the form
\begin{equation} \label{eq:NS3}
	0 \to \ZZ_4 \to \ZZ_4\oplus\ZZ^k \to \ZZ^k \to \ZZ_2 \to 0.
\end{equation}

\begin{figure}[htb!]
	\centering
	\subcaptionbox{$[\tilde{\ell}_z] = [-\tilde{\ell}_z]$\label{subfig:Z4_inverse}} {\includegraphics[width=.31\textwidth]{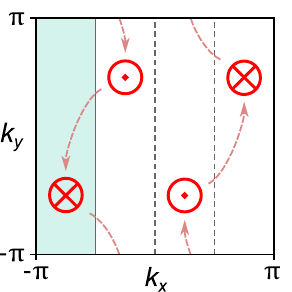}}
	\hfil
	\subcaptionbox{$[\tilde{\ell}_z] = 2g$\label{subfig:Z4_combine}} {\includegraphics[width=.31\textwidth]{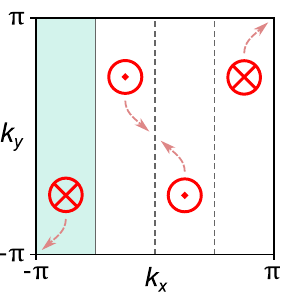}}
	\hfil
	\subcaptionbox{$g$\label{subfig:Z4_generator}} {\includegraphics[width=.31\textwidth]{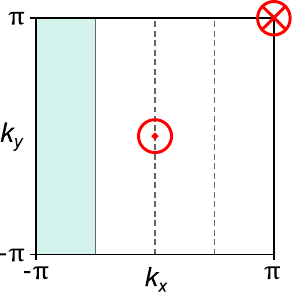}}
	\caption{Illustration of the insulating $\ZZ_4$ invariant of systems subject to $Pna2_1$ symmetry. The Dirac loop $\tilde{\ell}_z$ shown in (a) and (b) is the result of letting the symmetry act on a basic loop $\ell_z$ in $\TT^3$, keeping in mind the sign changes induced by the twisted homology: a loop running in the negative $k_z$-direction in the shaded fundamental domain is taken to three other loops in other quadrants of the Brillouin torus. The resulting element $\tilde{\ell}_z$ of the twisted homology group $H_1(M,\widetilde{\ZZ})$ is shown to be its own inverse in (a), indicating that it is an element of order two: the configuration of loops can be continuously deformed while respecting the symmetry, such that the loops end up changing directions. On the other hand, the same element is shown in (b) to be equal to twice a different element $g$: the four loops can be symmetrically moved to coincide with high-symmetry lines of the screw rotation in doubled pairs. Halving these doubled pairs results in the element $g$ shown in (c), which must be of order four, i.e. it acts as a generator of the $\ZZ_4$ group.}
	\label{fig:Z4_invariant} 
\end{figure}

The non-orientable Brillouin platycosms induced by all of the above space group symmetries naturally obey the same mod 2 charge cancellation theorem as $\KS$, as can be seen from similar features of the exact sequences: the final group of total charge is $\ZZ_2$ in each case, and the second group of semimetal invariants has a factor of $\ZZ^k$ rather than $\ZZ^{k-1}$ relating to Weyl point charges. 
The difference between the various space groups instead manifests itself via the group of insulating invariants, reflecting the fact that the reduced Brillouin zones have fundamentally different topologies under each symmetry.
Standing out among these is the $\ZZ_4$ insulating invariant in the $Pna2_1$ system, which is also described in Ref.~\cite{Zhang2025} (and more implicitly in Ref.~\cite{Shiozaki2022}, which provides a more technical treatment of the 230 space group symmetries in terms of spectral sequences in K-theory).
Ref.~\cite{Zhang2025} provides a concrete integral expression for this invariant. 
Rather than repeating this expression here, we note that physical intuition about this invariant can be drawn from the twisted homology picture, similar to the reasoning in Fig~\ref{fig:K2S1_invariants}. 
That is, the generator of the $\ZZ_4$ invariant can be represented by a configuration of twisted Dirac loops; this is illustrated in Fig.~\ref{fig:Z4_invariant}. 
In particular, creating and annihilating Weyl nodes along these twisted Dirac loops induces non-trivial $\ZZ_4$ topology.

\subsection{Non-Hermitian systems} \label{sec:NH}
As recent years have seen a vastly increased focus on the topology of operators in the non-Hermitian regime, paving the way to the field of non-Hermitian topological physics~\cite{NHreview}, we would be remiss not to include a discussion of the way in which our coordinate-free topological classification extends to these systems.
In non-Hermitian systems, the conventional Weyl points are replaced by so-called exceptional points, marking degeneracies at which not only the eigenvalues (i.e.\ the energy bands), but also the eigenvectors (i.e.\ the eigenstates of the Hamiltonian) coalesce~\cite{Kato}.
Remarkably, the generic appearance of these exceptional points requires the tuning of only two parameters; as a result, they may already appear as stable points in two dimensions, in contrast to Weyl points which require three dimensions to appear generically.
Non-Hermitian systems subject to orientation-reversing symmetries were recently studied in Refs.~\cite{Konig2025,Rui2025}, where the authors show that some of the conclusions valid in Hermitian systems carry over into the non-Hermitian regime.
However, the problems arising from the lack of orientability remain, once again motivating the need for an extended and coordinate-free topological classification of these systems.

Fortunately, certain important classes of non-Hermitian systems can be mapped canonically onto corresponding Hermitian semimetals, allowing the topology to be studied directly using the tools developed above.
The key concepts at work in this correspondence are the so-called \emph{resultant winding number}~\cite{Delplace21,Tang2023,Yoshida2024} and \emph{resultant Hamiltonian}~\cite{Stalhammar2025}, which can be employed to relate the topology of non-Hermitian exceptional points to that of Hermitian nodal points.
The most important concepts will be repeated here, but the reader is referred to Refs.~\cite{Yoshida2024,Stalhammar2025} for a more detailed background on this matter.
In the rest of this section, we employ a two-dimensional glide symmetry as an illustrative example.

A non-Hermitian Hamiltonian subject to a two-dimensional glide symmetry is bound to satisfy
\begin{equation}
    \Ham\left(k_x,k_y\right) = U\Ham\left(-k_x,k_y+\pi\right)U^{-1}.
\end{equation}
This Hamiltonian may generically host exceptional points, the location of which can be derived from its characteristic polynomial: in general, specific resultants of this polynomial and its derivatives will be zero at such points.
The characteristic polynomial is given by $P_2\left(\lambda;\bk\right) = \det\left[ \Ham\left(\bk\right)-\lambda I\right]$, and the corresponding resultant is given by
\begin{align}
    R(\bk) = \Res\left[P_2(\lambda;\bk),\partial_{\lambda}P_2(\lambda;\bk)\right].
\end{align}
This resultant is complex-valued in general, meaning it gives rise to a well-defined vector field on the two-dimensional Brillouin zone: 
\begin{equation}
    \mathbf{R}(\bk) = \left\{\Real\left[R(\bk)\right],\Imag\left[R(\bk)\right]\right\}.
\end{equation}
The zeros of $\mathbf{R}$ then correspond exactly to the position of the exceptional points~\cite{Yoshida2024,Stalhammar2025}.
Importantly, $\mathbf{R}(\bk)$ satisfies the same two-dimensional glide symmetry as the parent Hamiltonian, meaning that the topology of the zeros of $\mathbf{R}(\bk)$ (and hence the topology of the exceptional points of the parent Hamiltonian) can be described using a Mayer--Vietoris sequence coinciding with Eq.~\eqref{eq:K2-surface-cohomology}.
Therefore, the same mod 2 charge cancellation theorem readily holds for exceptional points in non-Hermitian systems in two dimensions, as well as the corresponding conclusions regarding different choices of the fundamental domain.
The conclusions regarding the $K^2$ surface Fermi arcs stemming from Sec.~\ref{sec:k2surface} are instead carried over to a completely analogous interpretation of the uniquely non-Hermitian bulk Fermi arcs, which are branch-cut like curves connecting exceptional points.

One fundamental limitation of this formalism should be pointed out: due to the commutative nature of (co)homology groups, it only captures Abelian topology in non-Hermitian systems, whereas such systems may host non-Abelian topology in general.
To be specific, this Abelian topology arises in the context of $n$-fold exceptional points emerging in $n$-band systems; for example, the Hamiltonian discussed above therefore needs to describe a two-band system, as exceptional points generically emerging in two dimensions are two-fold degenerate~\cite{NHreview}.

\subsection{Inversion symmetry} \label{sec:invsym}
As a last extension, we will outline a heuristic approach to the (co)homological classification of inversion-symmetric Weyl semimetals.
This type of Weyl semimetal is of marked experimental importance, since a material needs to break either time reversal or inversion symmetry in order to host Weyl nodes~\cite{weylreview}, but not necessarily both. Given that time reversal symmetry is naturally broken in the presence of a magnetic field, some prominent experimental realisations of Weyl semimetals have come in the form of inversion-symmetric \emph{magnetic Weyl semimetals}~\cite{Liu2019,Belopolski2019,Morali2019}.

At the level of the momentum-space Hamiltonian, inversion symmetry imposes the constraint 
\begin{equation}
    \Ham\left(\bk\right) = U_{\mathcal{I}}\Ham\left(-\bk\right)U^{-1}_{\mathcal{I}}.
\end{equation}
Note that this symmetry has an action in momentum space that is similar to time reversal symmetry, the key difference being that the latter includes an anti-unitary rather than unitary conjugation.
Importantly, inversion symmetry relates the same momenta on the Brillouin zone $\TT^3$ to each other; in particular, it induces the same set of eight fixed points, referred to as the time-reversal invariant momenta (TRIM).
As explained in Sec.~\ref{sec:motivation}, the fixed points resulting from a non-free symmetry action need to be treated with additional scrutiny in the corresponding (co)homology description.
For example, the classification of time-reversal symmetric Weyl semimetals in Ref.~\cite{Thiang2017} relies on the use of equivariant homology \emph{excluding} the TRIM, and twisted equivariant cohomology \emph{relative to} the TRIM. Here, the twist in the cohomology group and treatment of the TRIM are based on precise arguments related to the classification of special equivariant vector bundle structures over the Brillouin zone~\cite{Nittis2018}.
As noted in Sec.~\ref{sec:motivation}, the twist in cohomology agrees in particular with the fact that each Weyl node has a symmetric partner of the same chirality.

For inversion symmetry, however, the symmetric partners of Weyl nodes instead have the opposite chirality, due to its unitary orientation-reversing nature.
As such, the (co)homology classification of inversion-symmetric Weyl semimetals should rely on some form of twisted equivariant homology and ordinary equivariant cohomology.
What remains is to find the correct treatment of the behaviour around the TRIM, which is qualitatively very different from the time-reversal symmetric case: since a symmetric pair of Weyl nodes has opposite charges, they may be brought together and annihilated at any of the TRIM.
This leads to the possibility of a configuration with two Weyl points in the Brillouin torus, connected by a Dirac string running through one of the TRIM; such a configuration is impossible in a time-reversal symmetric Weyl semimetal, which must host at least four Weyl points~\cite{weylreview}.
In principle, similarly rigorous arguments based on vector bundle classification can be constructed for a proper treatment of these fixed points; while such a description is certainly desirable in its own right, it is beyond the scope of the present work.
Instead, we will demonstrate that the complete (co)homology description can be equally well obtained from a heuristic argument based on the aforementioned behaviour of Weyl points and Dirac strings at the TRIM.

\begin{figure}[htb!]
	\centering
	\subcaptionbox{\label{subfig:inversion_four}}{\includegraphics[width=.3\linewidth]{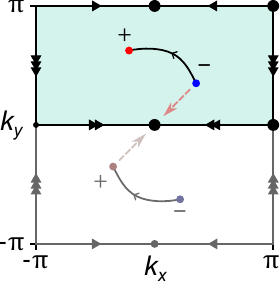}}
	\hfil
	\subcaptionbox{\label{subfig:inversion_two}}{\includegraphics[width=.3\linewidth]{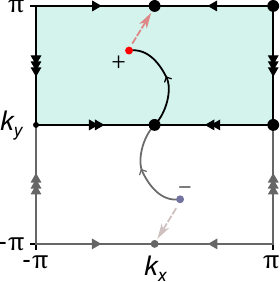}}
	\hfil
	\subcaptionbox{\label{subfig:inversion_zero}}{\includegraphics[width=.3\linewidth]{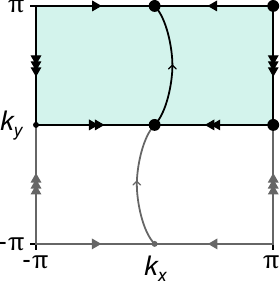}}
	\caption{Illustration of a two-dimensional slice of the Brillouin torus at $k_z=0$, with a choice of fundamental domain $M$ highlighted in teal. The TRIM are indicated by bold black dots, and arrows indicate the correct boundary identifications on $M$. Starting with a basic Dirac string connecting two Weyl points in $M$, a series of Weyl point annihilation is played out on the torus, while the apparent topology of (twisted) Dirac strings and loops is studied in isolation within the fundamental domain; the complementary area is greyed out as a reminder of this.
    (a) The Dirac string contained within the fundamental domain first appears similar to an element of $H_1(M,W;\ZZtwist)$. From this initial position, the negative Weyl point and its symmetric partner are brought together along the dashed arrows.
    (b) Eventually, these two Weyl points are annihilated at the TRIM at $\vb{k}=0$. From the perspective of the fundamental domain, it seems as though the negative Weyl point is absorbed by the TRIM. Consequently, the Dirac string appears to be bounded by this TRIM as well as at the remaining Weyl point; that is, a Dirac string on $M$ more properly acts as an element of $H_1\left(M,W\cup\text{TRIM};\ZZtwist\right)$. Note that this is still a valid description for panel (a), since the Dirac string depicted there terminates at $W\subset(W\cup\text{TRIM})$.
    (c) Annihilating the remaining two Weyl points at the TRIM at $\vb{k}=(0,\pm\pi,0)$ results in a Dirac loop that, from the perspective of the fundamental domain, runs between two different TRIM; in other words, Dirac loops act like elements of $H_1\left(M,\text{TRIM};\ZZtwist\right)$ in this system.}
	\label{fig:inversion_Dirac-strings}
\end{figure}

The idea is as follows: in all of the cases studied above, the free action of the symmetry group $G$ ensures that the fundamental domain takes on the topology of $\TT^3/G$ -- for example, $\KS\cong\TT^3/\ZZ_2$, where $\ZZ_2$ represents the single glide symmetry.
This allows the equivariant (co)homology on $\TT^3$ to be reduced to ordinary (co)homology on the quotient space.
Under the $\ZZ_2$ inversion symmetry, the fundamental domain $M$ ``almost'' has the topology of $\TT^3/\ZZ_2$; that is, all momenta in $M$ act like the quotient of two points on the torus, except the eight TRIM.
In light of this, Dirac strings in $M$ that do not interact with the TRIM essentially act like elements of the usual (non-equivariant) twisted homology group $H_1(M,W;\ZZtwist)$, where $W$ is the set of Weyl points in $M$.
Crucially, the aforementioned annihilation of one of the TRIM creates what seems from the perspective of $M$ to be a Dirac string which terminates at the TRIM; this argument is illustrated in detail in Fig.~\ref{fig:inversion_Dirac-strings}.
Recalling that homology relative to the set of Weyl points $W$ encodes the fact that Dirac strings are allowed to terminate at these points, this leads us to propose that the correct classification in the case of inversion symmetry is given by twisted homology on $M$ relative to the TRIM.
This leads to the following ansatz for the exact twisted homology sequence:
\begin{align}
	0\ &\to\ H_1(M, \text{TRIM}; \ZZtwist)\ \to\ H_1(M, W\cup\text{TRIM}; \ZZtwist)\ \nonumber \\
	&\quad \overset{\partial}{\to}\ H_0(W; \ZZtwist)\ \to\ H_0(M, \text{TRIM}; \ZZtwist)\ \to\ 0.
\end{align}
Lefschetz duality ensures that there is a corresponding Mayer--Vietoris sequence in ordinary cohomology \emph{excluding} the TRIM, which reads
\begin{align} \label{eq:inversion-MV}
	0\ &\to\ H^2\big(M\setminus\text{TRIM}\big)\ \to\ H^2\big(M\setminus W\cup\text{TRIM}\big)\ \nonumber \\
	&\quad\to\ \bigoplus_{w\in W} H^2(S_w^2)\ \to\ H^3\big(M\setminus\text{TRIM}\big)\ \to\ 0.
\end{align}
Incidentally, the exclusion of the TRIM from these cohomology groups ensures that they are once again completely equivalent to equivariant cohomology on $\TT^3\setminus\text{TRIM}$, so that the sequence above may also be conceptualised in those terms.

\begin{figure}[htb!]
	\centering
	\subcaptionbox{\label{subfig:inversion_EBZ}}{\includegraphics[width=.3\linewidth]{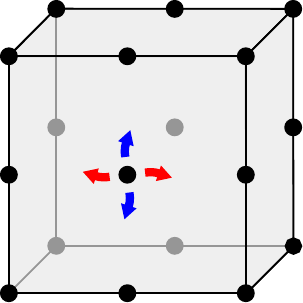}}
	\hfil
	\subcaptionbox{\label{subfig:inversion_retract}}{\includegraphics[width=.3\linewidth]{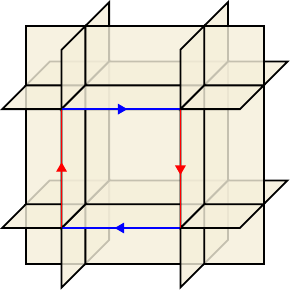}}
	\hfil
	\subcaptionbox{\label{subfig:inversion_translation}}{\includegraphics[width=.3\linewidth]{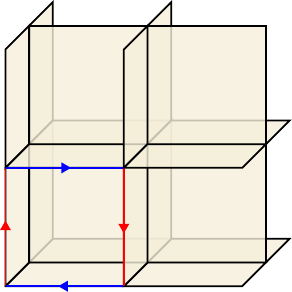}}
	\caption{Illustration of the process to obtain a cell structure for calculation of the inversion symmetric invariants. The effective Brillouin zone of the system is shown in (a) as the front half of the torus, where the front and back face are both individually subject to internal symmetric identifications. This volume is punctured for the calculation: the TRIM (black dots) are removed, leaving a non-compact space. The resulting holes can be ``grown'' symmetrically while maintaining the topology of the space, as indicated by the outgoing blue and red arrows. Doing this for all TRIM at once results in the set of two-dimensional planes shown in (b), with example boundary identifications shown around one of the resulting holes. In principle, this space can be used directly as the 2-skeleton of a cell structure; however, this cell structure can be simplified somewhat by performing a translation by $\pi/2$ in the two periodic directions, resulting in the structure depicted in (c).}
	\label{fig:inversion_calculation}
\end{figure}

Explicit calculation of this sequence is somewhat more involved than in the previously discussed cases, since $M\setminus\text{TRIM}$ is non-compact and does not immediately admit a cell structure.
To remedy this, a deformation retraction can be performed onto a suitable 2-skeleton; that is, the eight holes corresponding to the TRIM can be ``grown'' while maintaining the correct boundary identifications, until only a set of intersecting two-dimensional surfaces remains. 
This process is illustrated in Fig.~\ref{fig:inversion_calculation}.
The resulting cell structure is fairly complicated, and the necessary cellular homology calculations are aided by computational tools such as the Smith normal form~\cite{Peltier2005}.
Going through these calculations yields the following explicit expression for the Mayer--Vietoris sequence in Eq.~\eqref{eq:inversion-MV}:
\begin{equation}\label{eq:inversion-MV-explicit}
	0 \to \ZZ^3\oplus\ZZ_2^4 \to \ZZ^3\oplus\ZZ_2^4\oplus\ZZ^r \to \ZZ^r \to 0 \to 0,
\end{equation}
where $r$ denotes the number of Weyl points in the fundamental domain, i.e.\ the number of symmetric pairs of Weyl nodes on the torus.

The sequence in Eq.~\eqref{eq:inversion-MV-explicit} has one feature which immediately stands out: the total charge group appearing on the right is now the trivial zero group, whereas this group is $\ZZ$ on the basic torus and $\ZZ_2$ on each non-orientable Brillouin zone discussed previously.
Physically, this means that the notion of charge cancellation is completely absent on the fundamental domain -- here, the total chirality can take on any value.
Intuitively speaking, this relates to the fact that the system admits configurations such as those illustrated in Fig.~\ref{fig:inversion_Dirac-strings}(b), where the fundamental domain hosts a single Weyl point of unit charge.
It should be emphasised that the lack of charge cancellation on the fundamental domain is not an indication of any physical anomalies, just as in the case of $\KS$.
Instead, it should be understood in terms of the Weyl points being related to their own charge cancelling partner through the symmetry.

The $\ZZ^3\oplus\ZZ_2^4$ group of insulating invariants appearing on the left side of Eq.~\eqref{eq:inversion-MV-explicit} is most readily interpreted in terms of twisted Dirac loops, as was done for the glide symmetry in Sec.~\ref{sec:TIs}.
We take Fig.~\ref{fig:inversion_Dirac-strings}(c) as starting point, where an inversion-symmetric twisted Dirac loop is depicted which runs through the two TRIM at $\bk=(0,0,0)$ and $\bk=(0,\pi,0)$ along the positive $k_y$-direction.
We will denote this loop by $\ell_{0y0}$, where the three subscripts correspond to the three coordinate directions: the appearance of a zero in the $k_x$ and $k_z$ positions indicates that the loop sits at fixed $k_x=k_z=0$, while the $y$ indicates that it is oriented in the positive $k_y$-direction.
This loop cannot be removed from the two TRIM it crosses by symmetry-preserving continuous deformations, meaning it is topologically distinct from the three other possible Dirac loops $\ell_{0y\pi}$, $\ell_{\pi y0}$, and $\ell_{\pi y\pi}$ that run through two TRIM in the positive $k_y$-direction (for instance, $\ell_{0y\pi}$ runs through the two TRIM at fixed $k_x=0$ and $k_z=\pi$).
Taken together with the four similarly defined loops in the $k_x$ and $k_z$-directions (denoted $\ell_{x00}$, $\ell_{x0\pi}$, $\ell_{x\pi 0}$, $\ell_{x\pi \pi}$ and $\ell_{00z}$, $\ell_{0\pi z}$, $\ell_{\pi 0 z}$, $\ell_{\pi \pi z}$, respectively), there are a total of 12 such basic twisted Dirac loops.
The twisted homology classes of these 12 loops generate the $\ZZ^3\oplus \ZZ_2^4$ group; this can be understood more precisely by studying the ways in which they can be detached from the TRIM points.

\begin{figure}[htb!]
	\centering
	\subcaptionbox{\label{subfig:inversion_detach}}{\includegraphics[width=.35\linewidth]{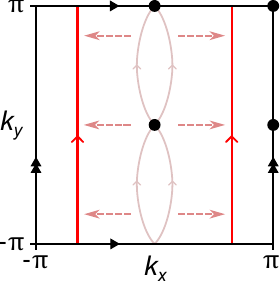}}
	\hfil
	\subcaptionbox{\label{subfig:inversion_equiv}}{\includegraphics[width=.35\linewidth]{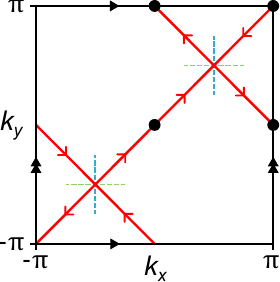}}
	\caption{(a) Two copies of $\ell_{0y0}$ can be moved apart in opposite directions while respecting inversion symmetry. Moving the loops even further apart makes them intersect the TRIM at $k_x=\pi$ simultaneously, proving that in terms of twisted homology classes, $2[\ell_{0y0}] = 2[\ell_{\pi y0}]$. (b) Intermediate state between $\ell_{0y0} - \ell_{\pi y0}$ and $\ell_{x00} - \ell_{x\pi0}$. Detaching the intersections along the vertical blue dashed lines (as though cutting them with scissors) yields $\ell_{0y0} - \ell_{\pi y0}$, whereas cutting along the horizontal green lines gives $\ell_{x00} - \ell_{x\pi0}$; here, the negative sign in both expressions is reflected in the fact that the corresponding loops run in the negative coordinate direction.}
	\label{fig:inversion_relations}
\end{figure}

As an example of this, consider a situation in which two loops of the form $\ell_{0y0}$ are present at the same time.
If two such loops are present, they can be detached from $(0,0,0)$ and $(0,\pi,0)$ in a symmetry-preserving fashion; this is illustrated in Fig.~\ref{fig:inversion_relations} (a).
After moving the two loops apart, they can instead be made to simultaneously intersect the TRIM at $(\pi,0,0)$ and $(\pi,\pi,0)$, meaning that they have been deformed into two copies of $\ell_{\pi y0}$.
As a result, the twisted homology classes $[\ell_{0y0}]$ and $[\ell_{\pi y0}]$ are interdependent, and therefore unable to generate separate $\ZZ$ invariants; this is precisely why only three factors of $\ZZ$ appear in the invariant group, corresponding to the three coordinate directions.
Meanwhile, the difference between these two classes is necessarily a $\ZZ_2$ invariant since
\begin{equation}
    2[\ell_{0y0}]=2[\ell_{\pi y0}] \implies 2[\ell_{0y0}-\ell_{\pi y0}]=0.
\end{equation}
Furthermore, a simple argument depicted in Fig.~\ref{fig:inversion_relations}(b) shows that this difference $[\ell_{0y0}-\ell_{\pi y0}]$ is equivalent to the difference $[\ell_{x00}-\ell_{x\pi0}]$ of two loops running in the positive $x$ direction.

Functionally similar arguments give rise to additional relations between the basic twisted Dirac loops. All in all, we conclude that the insulating invariant group $\ZZ^3\oplus\ZZ_2^4$ admits a basis of three $\ZZ$ generators given by
\begin{equation}
    \nu_x \coloneqq [\ell_{x00}], \quad \nu_y\coloneqq[\ell_{0y0}], \quad \nu_z\coloneqq[\ell_{00z}],
\end{equation}
along with four $\ZZ_2$ generators given by
\begin{alignat}{2}\label{eq:Z2-generators}
    \nu_{xy} &\coloneqq [\ell_{x00}-\ell_{x\pi 0}]=[\ell_{0y0}-\ell_{\pi y0}], \qquad \nu_{xyz} &&\coloneqq [\ell_{x00}-\ell_{x\pi 0}-\ell_{x0\pi}+\ell_{x\pi \pi}]\nonumber
    \\
    \nu_{yz} &\coloneqq[\ell_{0y0}-\ell_{0y\pi}] = [\ell_{00z}-\ell_{0\pi z}], &&\noncoloneqq [\ell_{0y0}-\ell_{\pi y0}-\ell_{0y\pi}+\ell_{\pi y\pi}]
    \\
    \nu_{xz} &\coloneqq [\ell_{00z}-\ell_{\pi 0 z}] = [\ell_{x00}-\ell_{x0\pi}], &&\noncoloneqq [\ell_{00z}-\ell_{\pi 0z}-\ell_{0\pi z}+\ell_{\pi \pi z}], \nonumber
\end{alignat}
all together generating the full topological description of the insulating phases.

As before, the semimetal invariants in the group $\ZZ^3\oplus\ZZ_2^4\oplus\ZZ^k$ can be fully understood in terms of this insulating description, in the sense that the creation and annihilation of pairs of Weyl nodes along the loops discussed above mediates phase transitions between different insulating invariants.

We finish this section by noting that our description in terms of twisted Dirac loops agrees well with previously established conclusions in the context of inversion-symmetric insulators.
In particular, a many-band description of such insulators is given in Ref.~\cite{Turner2012}, in terms of three $\ZZ$-valued Chern numbers and eight other $\ZZ$ invariants based on eigenvalues of the inversion symmetry at each of the eight TRIM.
In the present two-band setting, the latter eight $\ZZ$ invariants are reduced to $\ZZ_2$ invariants, corresponding to the $\ZZ_2$ symmetry eigenvalue on the occupied band at each TRIM.
These eight are further reduced to the four independent $\ZZ_2$ factors appearing in the group $\ZZ^3\oplus\ZZ_2^4$ by imposing four physical constraints on these occupied states: concretely, the eigenvalues $\zeta\in\ZZ_2\coloneqq\{\pm1\}$ are bound to satisfy
\begin{equation}
    \prod_{\kappa\in \text{TRIM}}\zeta(\kappa) = 1, \quad \prod_{\kappa\in \text{TRIM}_{k_i=0}} \zeta(\kappa) = (-1)^{C_i}
\end{equation}
for insulating states. 
Here $\text{TRIM}_{k_i=0}$ denotes the four TRIM lying in the $k_i=0$ plane for any $i\in\{x,y,z\}$, and $C_i$ is the corresponding Chern number as calculated in Eq.~\eqref{eq:insulatorchern}. 
These four constraints dictate how the eight $\zeta\in\ZZ^2$ are interrelated, leaving only four factors effectively independent.
This has a precise interpretation in terms of the homology framework given above: the eigenvalue at any given TRIM is $-1$ precisely when a twisted Dirac loop or string passes through that point. 
The four independent combinations of eigenvalues can then be deduced from the Dirac loops in the four $\ZZ_2$ generators in Eq.~\eqref{eq:Z2-generators}: for example, $\nu_{xy}$ features Dirac loops crossing the four TRIM in the $xy$-plane, and as such corresponds to a state with an eigenvalue of $-1$ at those points and 1 elsewhere.

\section{Summary, discussion and outlook} \label{sec:sumdisout}

\subsection{Summary} \label{sec:summary}
In this work we have provided a coordinate-free description classifying the topology of Weyl semimetals with non-orientable Brillouin zones arising from orientation-reversing symmetries.
We show that the topological features induced by such symmetries can be understood within a framework employing certain exact sequences in (co)homology.
Rather than admitting a description in terms of ordinary homology and cohomology groups, the loss of orientability breaks conventional Poincar\'e duality, which invalidates the standard correspondence between the cohomology and homology descriptions.
This Poincar\'e duality is restored by introducing a twist in either the homology or the cohomology groups.
Through a straightforward argument involving the impact of orientation reversal on the Weyl points and their concomitant Dirac strings, we show that the correct choice of this twist can be arrived at by studying pairs of Weyl points related by orientation-reversing symmetry: opposite chiralities in such a pair indicate the use of twisted homology, whereas pairs of the same chirality indicate that the cohomology is twisted.

This framework is then applied to the case of a $\KS$ Brillouin zone arising from a single momentum-space glide symmetry.
Identifying the correct Mayer--Vietoris sequence in ordinary cohomology groups and its dual sequence in twisted homology, we provide a derivation of the mod 2 charge cancellation theorem existing in this context, and classify the accompanying topological invariants for the insulating and semimetallic phases.
Importantly, the coordinate-free description demonstrates that any apparent non-zero net chirality in the $\KS$ fundamental domain is unphysical in nature, being dependent on an arbitrary choice of basis for the group of local charges; physically, this corresponds to a given position of the fundamental domain within the torus.
The topological invariants of the system are interpreted both in terms of integration of (cohomological) differential forms, and in terms of (homological) twisted Dirac strings, and the corresponding surface states on $K^2$-like surfaces are derived through the introduction of a projection map and its corresponding pushforward on homology.

The classification scheme is then applied in three additional contexts.
First, we complete the classification of all possible three-dimensional non-orientable Brillouin zones, by studying all orientation-reversing space group symmetries with a free action.
Four such topologies are found to exist in total: the previously studied $\KS$ arising from the space group $Pc$, and three other topologies related to the groups $Cc$, $Pca2_1$, and $Pna2_1$.
In all cases, similar mod 2 charge cancellation conditions are derived, but the insulating and semimetallic invariants are shown to differ between these groups.
Second, the exact sequences on $K^2$-like surfaces are related to non-Hermitian systems, and used to elucidate the mod 2 charge cancellation for exceptional points in non-Hermitian two-band systems.
Finally, we use heuristic arguments to provide a topological classification of inversion-symmetric Weyl semimetals in terms of twisted homology groups relative to the TRIM points. This reproduces the known insulating invariants and reframes them in terms of twisted Dirac loops, while at the same time providing a sensible understanding of the topology of the Weyl points; together, this allows the full semimetallic topology to be readily understood.

\subsection{Discussion and outlook} \label{sec:discussion}
The topological classification scheme in terms of (co)homology groups developed in this work not only serves as a fundamental theoretical framework encoding theoretical aspects of topological invariants and their origins, but also proves to be remarkably powerful from the perspective of physical interpretation.

One of the principal aims of this work has been to clarify the physical interpretation of the mod 2 charge cancellation theorem present on non-orientable Brillouin zones.
Our conclusion in this regard is that no novel physical phenomena are associated with this modified charge cancellation, and in particular the apparent non-zero total chirality which may arise in a chosen reduced Brillouin zone does not indicate any form of chiral anomaly. 
Instead, such non-zero total chirality is symptomatic of the ambiguities introduced by the choice of a single non-orientable fundamental domain.
The more physically conservative interpretation arising from our (co)homology treatment is that the relative charge between distinct Weyl points in the reduced Brillouin zone is ill-defined; because of this, the total charge is only well defined as a $\ZZ_2$ parity, and it is confined to be zero as usual.
Choosing a particular fundamental domain fixes these relative chiralities, leading to the appearance of non-zero total charge.
In this sense, there is a minor conceptual analogy with the ghost fields of quantum field theory, which arise when certain unphysical gauge degrees of freedom are fixed. These fields may manifest as ghost particles appearing in valid Feynman diagrams; still, these ghost particles can be removed by a proper change of gauge, meaning they are purely mathematical in nature.
Similarly, the net non-zero Weyl point chirality appearing in a chosen fundamental domain can be altered by repositioning the fundamental domain.

Importantly, the lack of phenomenological consequences does not necessarily mean that the classification scheme developed in this work is of no experimental relevance.
On the contrary, it provides a clear interpretation of the results of already conducted experiments~\cite{fonseca2024}, as well as any potential future experiments.
In particular, the homology interpretation of Dirac strings and Fermi arcs provides a solid mathematical and physical understanding of the global structure of systems such as the one studied in Ref.~\cite{fonseca2024}.
In this regard, we emphasise that we do not question the validity of the actual measurements performed in that work. Rather, we want to convey the opinion that observations of this nature should be accompanied by complementary analysis on the full Brillouin torus, where the phenomenology is more transparent; in particular, the appearance of surface Fermi arcs connecting equal chirality Weyl nodes is resolved in this context.
In this sense, we consider the reduced Brillouin zone to be mainly of theoretical interest, being the minimal space on which the complete physics of the system is captured. Correctly interpreting this physics relies on making correct global extrapolations, especially when experimental observations are involved.
This point is reinforced by comparison to inversion symmetry, where using the full Brillouin torus is standard practice: we demonstrate in Sec.~\ref{sec:invsym} that an effective Brillouin zone can be used here with enough care, and an ordinary configuration such as that in Fig.~\ref{subfig:inversion_two} looks unconventional from this frame of reference.

As far as experiments go, one interesting potential avenue for future observations lies in a fully three-dimensional realisation of non-orientable Weyl semimetals in the context of acoustic crystals. 
Non-symmorphic momentum-space symmetries have been successfully studied in this context~\cite{Wang2023,Tao2024,Zhu2024}, as have Weyl semimetals~\cite{Xiao2020,Wang2021,Li2025}; combining these two aspects would allow many of the physical properties of the systems studied in this work to be probed directly.

Besides experimental considerations, a compelling potential physical implication also arises from our classification of insulating inversion-symmetric systems.
It is shown in Ref.~\cite{Turner2012} that many-band inversion-symmetric topological insulators can be divided into 16 different classes, distinguishable by their transport properties, for any given Chern vector.
This might be explained precisely by the $\ZZ^3\oplus\ZZ_2^4$ structure of the invariant group that we find in the two-band homology treatment: the Chern vector fixes an element of the first $\ZZ^3$ summand, while the remaining $\ZZ_2^4$ summand has precisely $2^4=16$ elements.
We have not made this correspondence precise, but in general it should not be surprising that transport properties are well described by two-band models: such models focus solely on dispersion at the Fermi level by isolating the valence and conduction bands.
If the correspondence does indeed turn out to be exact, this highlights the great potential of two-band descriptions in predicting physical features which are also present in multi-band models.

It should also be mentioned that circumventions of the Nielsen--Ninomiya theorem have been studied somewhat extensively in other setups, including Floquet systems~\cite{Sun2018,Higashikawa2019,Zhu2020} and flux-biased Weyl superconductors~\cite{Brien2017}, where the system setups seemingly do result in observable phenomena, such as equilibrium chiral magnetic effects sourced by a net chiral anomaly~\cite{Stalhammar2024}.
Whether or not these systems are even remotely related to the notion of non-orientable Brillouin zones, or how these are to be treated in terms of a classification scheme similar to the one developed in this work, comprises an interesting question to be answered in future research projects.

On a more theoretically fundamental level, there are several open questions that deserve to be addressed in future works.
One immediate such question originating from the current work concerns the physics of additional surface states on non-orientable Brillouin zones.
In this work, the surface states obtained by projecting out the periodic $S^1$ factor of $\KS$ are studied, but potential topological surface states are also bound to exist upon projection along the other two coordinate directions.
These other projections interact with the glide symmetry in a non-trivial way, and as such they do not necessarily preserve its orientation-reversing properties.
This leads us to believe that such surface states may need to be classified using a more involved and modified (co)homology description.
An additional avenue for research lies in the formulation of similar classifications for systems subject to other symmetries, both with free and non-free actions.
In particular, venturing into the realm of symmetries with non-free actions opens the possibility to provide similar (co)homology descriptions of systems subject to two mutually orthogonal glide symmetries, whose fundamental domains share some topological properties with the real projective space $\RR P^2$, or in three dimensions, its direct product with the circle $\RR P^2\times S^1$ -- notably, the presence of fixed momenta under this symmetry ensures that these spaces do not act as proper Brillouin zones.
Such fundamental domains are studied in Refs.~\cite{fonseca2024,Konig2025}, indicating their immediate interest in the community.
To the knowledge of the authors, descriptions in the specific terms of dual exact sequences for homology and cohomology exist for systems without symmetries~\cite{Mathai2017a,Mathai2017b} (a description that can be directly carried over to systems subject to chiral symmetry), and systems subject to time reversal symmetry~\cite{Thiang2017}.
Hermitian systems subject to the simultaneous combination of parity and time reversal symmetry, resulting in so-called knotted semimetals, have also been studied in this context, although a satisfactory classification scheme remains elusive to date~\cite{Celeste2023}.
It should be noted that descriptions of many of these symmetries have been provided in the the context of K-theory, which is a generalised cohomology theory; notably, Ref.~\cite{Shiozaki2022} provides a K-theoretic classification of systems subject to all 230 space groups, including semimetallic information.
However, these works generally do not exploit the relation between local and global topology which is contained in the Mayer--Vietoris sequences studied here, nor the Poincar\'e duality giving rise to dual homology descriptions, while both of these aspects prove to contribute much intuitive physical insight.

In general, there is vast potential for expansion into this non-Hermitian regime.
As previously mentioned, the case of two-dimensional glide-symmetric non-Hermitian systems and mod 2 charge cancellations of exceptional points has recently been studied in Refs.~\cite{Konig2025,Rui2025}.
One interesting observation made in Ref.~\cite{Konig2025} pertains to specific kinds of non-Abelian phase transitions related to the so-called rebasing operation.
Such transitions are sensitive to similar fundamental domain alterations as those discussed in this work, and as such may not represent true transitions between distinct topological phases of the system.
Possibly, the homotopy description of this phenomenon is better understood in terms of fundamental groupoids rather than fundamental groups, in which such rebasing operations induce isomorphisms between fundamental groups.
A proper (co)homology treatment of this behaviour is confounded by its non-Abelian nature. 

More generally, non-Hermitian systems have been extensively studied in terms of homotopy theory in recent years to account for these non-Abelian features~\cite{Wojcik2020,Li2021,Yang2024}. 
Such features are not exclusive to the non-Hermitian realm: they may also emerge within Hermitian band structures, for instance when considering symmetry-protected nodal lines rather than nodal points~\cite{Bzdusek2017,Sun2018a}. 
(Co)homology descriptions of these Hermitian systems naturally run into similar limitations with respect to commutativity, and as such they also necessitate complementary classifications using e.g.\ homotopy theory~\cite{Wu2019} or the use of different characteristic classes~\cite{Ahn2019}.

\section{Conclusion} \label{sec:conclusion}
The interplay between condensed matter physics and topology has revolutionised modern physics research in recent decades, paving the way towards state-of-the-art semiconductor devices and notable achievements in quantum information -- with semimetallic and superconducting implementations marking future milestones.
Many key material properties are readily described by the interplay between the valence and conduction bands, thus offering a compelling explanation even within the theory of two-band models.
Even so, fundamental topological classifications of these properties stemming from the framework of algebraic topology are surprisingly rare, despite their potential in providing not only a fundamental mathematical understanding of their origin, but also their physical interpretation.

Our work highlights how the complete understanding of topological semimetals and insulators benefits from such a fundamental topological description.
This is exemplified through the use of twisted (co)homology for the classification of semimetals subject to orientation-reversing symmetries, including space group symmetries with free actions and inversion symmetry.
In particular, our classification scheme provides a proper physical understanding of the accompanying modified cancellation theorems, and remedies the corresponding interpretation of net non-zero chiralities and chiral anomalies in these systems.
We also show that the classification scheme can be extended further to the non-Hermitian regime, giving it further relevance in the context of dissipative and open quantum systems and paving the way towards an increased physical and mathematical understanding of systems relevant in a wide range of different areas of today's physics research.

\section*{Acknowledgements}
Both authors express their gratitude to Cristiane Morais Smith for extensive discussions and supervision throughout the project.
Scientific and social discussions with Diego Felipe Mu{\~n}oz-Arboleda, Lumen Eek, Anouar Moustaj, David Santiago Quevedo Vega, and Robin Verstraten are further acknowledged.
MS is supported by the Swedish Research Council (VR) under grant number 2024.00272.

\bibliographystyle{SciPost_bibstyle}

\end{document}